%% file: 0_main.tex
\documentclass[letterpaper]{article} 
\usepackage{aaai25}  
\usepackage{times}  
\usepackage{helvet}  
\usepackage{courier}  
\usepackage{graphicx} 
\usepackage[hyphens]{url}  
\usepackage{graphicx} 
\usepackage{natbib}  
\usepackage{caption} 
\usepackage{algorithm}
\usepackage{algorithmic}
\usepackage{xr}
\usepackage{newfloat}
\usepackage{listings}
\DeclareCaptionStyle{ruled}{labelfont=normalfont,labelsep=colon,strut=off} 
\floatstyle{ruled}
\newfloat{listing}{tb}{lst}{}
\floatname{listing}{Listing}
\usepackage{booktabs}
\usepackage{enumitem}

\usepackage{multirow}
\usepackage{array}  
\newcolumntype{L}[1]{>{\raggedright\arraybackslash}p{#1}}
\usepackage[most]{tcolorbox}
\tcbuselibrary{breakable}

\title{The AI Model Risk Catalog: What Developers and Researchers Miss About Real-World AI Harms}
\author{
    Pooja S. B. Rao\textsuperscript{\rm 1,3},
    Sanja \v{S}\'{c}epanovi\'{c}\textsuperscript{\rm 2,4}, 
    Dinesh Babu Jayagopi\textsuperscript{\rm 3},
    Mauro Cherubini\textsuperscript{\rm 1},\\
    Daniele Quercia\textsuperscript{\rm 2,5}
}
\affiliations{
    \textsuperscript{\rm 1}University of Lausanne, Lausanne, Switzerland\\
    \textsuperscript{\rm 2}Nokia Bell Labs, Cambridge, UK\\
    \textsuperscript{\rm 3}International Institute of Information Technology Bangalore, Bangalore, India\\
    \textsuperscript{\rm 4}University of Oxford, Oxford, UK\\
    \textsuperscript{\rm 5}Politecnico di Torino, Turin, Italy\\
        \{pooja.rao, mauro.cherubini\}@unil.ch,
        jdinesh@iiitb.ac.in,
        \{sanja.scepanovic, daniele.quercia\}@nokia-bell-labs.com
}

\usepackage{xcolor}

\newcommand\revision[1]{\textcolor{black}{#1}}

\begin{document}

\maketitle

\begin{abstract}
We analyzed nearly 460,000 AI model cards from Hugging Face to examine how developers report risks. From these, we extracted around 3,000 unique risk mentions and built the \emph{AI Model Risk Catalog}. We compared these with risks identified by researchers in the MIT Risk Repository and with real-world incidents from the AI Incident Database. Developers focused on technical issues like bias and safety, while researchers emphasized broader social impacts. Both groups paid little attention to fraud and manipulation, which are common harms arising from how people interact with AI. Our findings show the need for clearer, structured risk reporting that helps developers think about human-interaction and systemic risks early in the design process. The catalog and paper appendix are available at: \url{https://social-dynamics.net/ai-risks/catalog}.

\end{abstract}

\input{sections/1_introduction}
\input{sections/2_related_work}
\input{sections/3_methods}
\input{sections/4_results}
\input{sections/5_results}
\input{sections/6_discussion}

\input{sections/statements}

\bibliography{manual, facct2025}

\input{appendix}

\end{document}

%% file: sections/1_introduction.tex
\section{Introduction}\label{sec:intro}
Risk and harm have been central concerns in AI safety and ethics research. Researchers have worked to identify and organize these concepts into taxonomies~\cite{weidinger2022taxonomy,yampolskiy2016taxonomy}. Some focus on Large Language Models and Generative AI~\cite{weidinger2021ethical,weidinger2022taxonomy,stahl2024ethics}, while others address broader AI systems~\cite{yampolskiy2016taxonomy,wirtz2020dark} or Artificial General Intelligence~\cite{mclean2023risks}.
%
These efforts often use different definitions of risk and harm. We follow the OECD~\cite{oecd2024aiincidents}, which defines risk as the chance of harm, and harm as a risk that has caused damage. In some contexts, the terms are used interchangeably. The MIT Risk Repository~\cite{slattery2024ai} compiles risks from academic work on AI frameworks and is the largest collection of researcher-identified AI risks to date. The AI Incident Database~\cite{mcgregor2021preventing} catalogs real-world harms, which have been further classified by \citet{velazquezDecodingRealWorldArtificial2024}.


Existing research has not explored how developers describe the risks of specific models or how those models might fail in typical user scenarios. This perspective is critical, as developers can offer grounded insight into the behavior and limitations of the systems they build.

\begin{figure*}[t]
    \centering
    \includegraphics[scale=0.27]{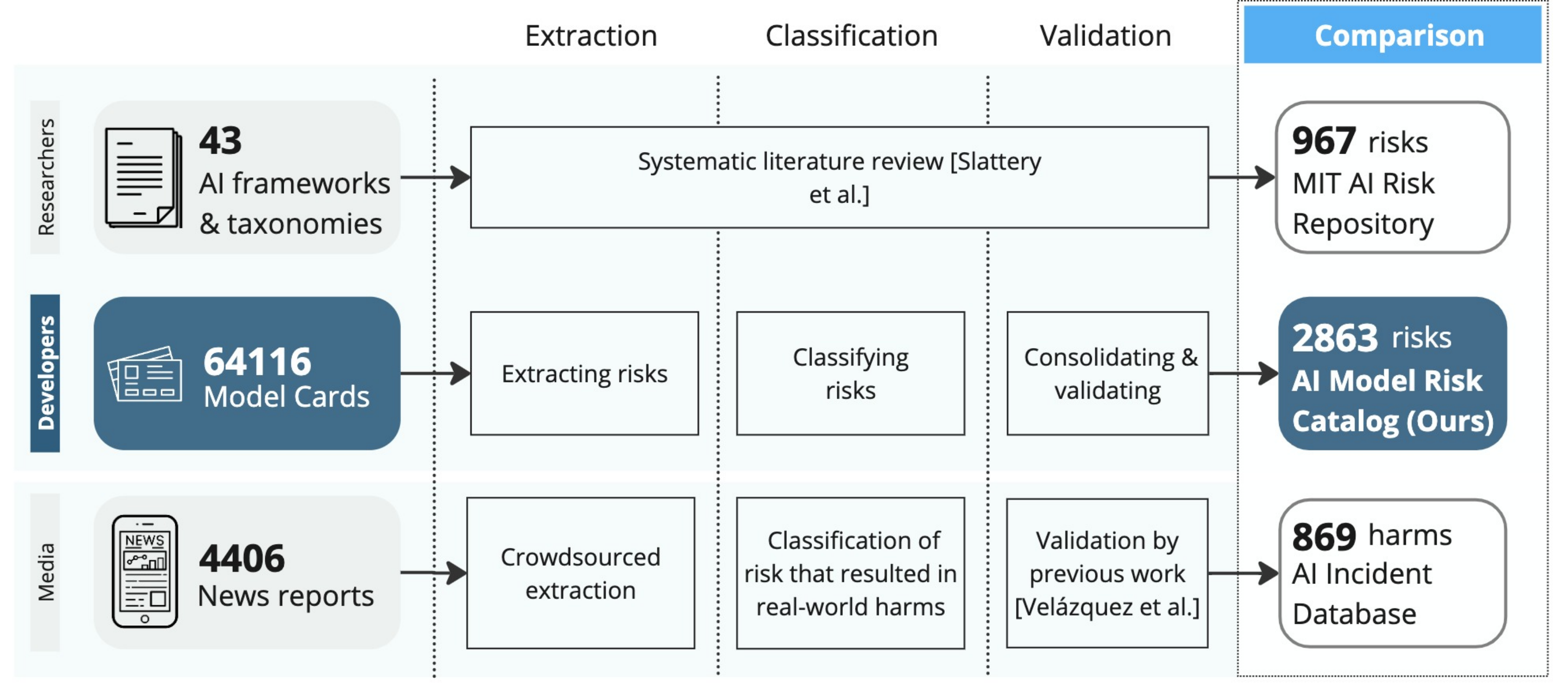}
    \caption{Overview of our methodology. Our methodology has three steps: extracting, classifying, and consolidating risks from model cards to build the AI Model Risk Catalog. We then compared the developer-reported risks in the catalog to those identified by researchers in the MIT Risk Repository~\cite{slattery2024ai} and to real-world harms from the AI Incident Database~\cite{mcgregor2021preventing}. The catalog is based on 64,116 model cards with risk sections, from which we extracted 37,401 risk mentions (including duplicates). After consolidation, we identified 2,863 unique risks from 2,672 model cards, making this the largest collection of AI risks across the three sources.
    }
    \label{fig:overview}
\end{figure*}


To address the lack of risk data linked to specific models and described by their developers, we analyzed nearly half a million model cards \revision{\cite{mitchellModelCardsModel2019}} from Hugging Face as of July 2024. \revision{Model cards are a widely adopted standard for documenting AI models, adopted by both major technology companies and individual developers.} This work makes two main contributions (Figure~\ref{fig:overview}):

\begin{enumerate}
\item \emph{Analyzing risks identified by developers to create the AI Model Risk Catalog.} We collected all available model cards on Hugging Face and 
using 
established taxonomies~\cite{weidinger2022taxonomy,slattery2024ai}, grouped similar risks within each category, and selected representative examples. The result is the AI Model Risk Catalog, which includes 2,863 categorized risks tied to specific models. The catalog is publicly available at: \url{https://social-dynamics.net/ai-risks/catalog}.


\item  \emph{Comparing developer and researcher risks with real-world harms.} We compared the types of risks emphasized in three sources: developer-described risks from model cards, researcher-identified risks from the MIT Risk Repository, and real-world harms from the AI Incident Database. Developers tend to report technical risks such as model limitations, safety issues, and bias. These account for over half the harms recorded in real-world incidents. Researchers focus on governance, societal impacts, and threats to human agency, which together explain fewer than 15\% of incidents. Notably, the largest share of incidents involves malicious use and misinformation. These risks  are underrepresented by both developers and researchers, likely because they depend on unpredictable human behavior: hard for researchers to anticipate, and even harder for developers to account for~\cite{velazquezDecodingRealWorldArtificial2024}.

    %
\end{enumerate}

We close by discussing how our catalog complements existing efforts to map AI risk and what it means for developers, researchers, journalists, policymakers, and the public.


%% file: sections/2_related_work.tex
\section{Related Work}\label{sec:rw}
We start by reviewing research on AI risk taxonomies, followed by that on collating AI risks into databases.

\subsection{AI risk taxonomies}\label{sec:rw_taxonomies}
%
The taxonomization of AI risks has grown significantly in recent years~\cite{yampolskiy2016taxonomy,liutrustworthy,stahl2024ethics,shelby2023sociotechnical}. 
One of the most widely used taxonomies proposed by DeepMind~\cite{weidinger2022taxonomy}, categorizes risks into six primary groups: \emph{representation and toxicity, misinformation, information and safety, malicious use, human autonomy and integrity,} and \emph{socioeconomic and environmental} harms. Each category is further divided into a total of 21 subcategories, such as \emph{toxic content} under the first category and \emph{environmental damage} under the last. \citet{weidinger2023sociotechnical} also explore the context of harm origins, identifying three layers: \emph{capability} (related to the system's technical features), \emph{human interaction} (linked to users' experiences with the system), and \emph{systemic impact} (including the broader context in which the system operates).

Other notable taxonomies include that by \citet{shelby2023sociotechnical}, which identifies five key harm categories: presentation, allocative, quality of service, interpersonal, and social system harms. \citet{yampolskiy2016taxonomy} proposes a classification based on the causes of risks, organized along three axes: entity (human or AI), intentionality (intentional or unintentional), and timing (pre-deployment or post-deployment).
Arguing that many existing taxonomies, including those mentioned above, are primarily designed for researchers and policymakers, making them less accessible to broader audience, \citet{abercrombie2024collaborative} introduced the AI, Algorithmic, and Automation Harms Taxonomy, which aims to be more comprehensible to the general public. This framework  adds categories such as physical and psychological harms. 
%


\subsection{AI risks data sources}\label{sec:rw_classifying_risks}
%
Another important area of research focuses on collating AI risks into data sources.

\paragraph{Risks envisioned by researchers.}  
In the most comprehensive effort to date, \citet{slattery2024ai} analyzed 43 scholarly and industry frameworks proposing various AI risk taxonomies and compiled them into the MIT AI Risk Repository. This continuously updated repository currently lists 967 risks, categorized using two taxonomies: a causal taxonomy adapted from \citet{yampolskiy2016taxonomy} and a domain taxonomy extended from the DeepMind taxonomy~\cite{weidinger2021ethical}. Using a best-fit approach~\cite{carroll2013best}, \citet{slattery2024ai} refined the DeepMind taxonomy by adding a seventh category, \emph{AI system safety, failures, and limitations}, and making minor adjustments to existing categories. 
%
 \citet{derczynski2023assessing} introduced ``Risk Cards'' providing definitions, categorizations, and examples of harms tied to specific language models (LMs). They identified risks from scholarly literature, categorized them using the taxonomies of \citet{weidinger2021ethical} and \citet{shelby2023sociotechnical}, and released these findings as the LM risk cards starter set. This set, however, is a result of a more specific effort (i.e., focuses on LMs only), and features significantly fewer risks compared to the MIT AI Risk Repository.

\paragraph{Automatically envisioned risks.}  
Tools have also been developed that often use LLMs to generate risk ideas and help AI practitioners anticipate risks during the design phase. For instance, \citet{herdel2024exploregen} created ExploreGen, an LLM-based tool that generates potential AI uses, and categorizes their risk level according to the EU AI Act. \citet{wang2024farsight} developed FarSight, an interactive tool that supports AI prototyping by providing news articles on related incidents and helping practitioners explore potential risks and affected stakeholders. \citet{buccinca2023aha} introduced AHA!, a tool leveraging vignettes to describe possible harms with the help of LLMs or crowdsourcing. CardGen \cite{liu2024automatic} is a pipeline that uses retrieval-augmented generation (RAG) to complete missing sections of model cards, including risk sections, by drawing on information from papers and GitHub projects. Lastly, RiskRAG is another RAG tool assisting developers in envisioning risks of their models that combines ExploreGen, and curates data from other model cards and the AI Incident database~\cite{rao2025riskrag}.

\paragraph{AI risks materialized as harms.}  
To track cases where AI systems have caused real-world harm \cite{buolamwini2018gender}, several databases of AI incident reports have been developed~\cite{rodrigues2023artificial}. The most well-known is the AI Incident Database~\cite{mcgregor2021preventing}, which curates news reports on AI failures and categorizes these incidents~\cite{turri2023we,abercrombie2024collaborative}. 

%
\citet{velazquezDecodingRealWorldArtificial2024} classified incidents recorded in the incident database using the DeepMind taxonomy~\cite{weidinger2022taxonomy}, finding that most incidents stemmed from the human-interaction layer, with harms most frequently falling under the human autonomy and integrity, or representation and toxicity categories. 
\citet{bogucka2024atlas} visualized data from the incident database, including a subset focused on mobile computing~\cite{bogucka2024mobileatlas}, making the information more accessible to broader audiences.
Other prominent databases include the AI, Algorithmic, and Automation Incidents and Controversies~\cite{pownall2023aiaaic}, the OECD Monitor~\cite{oecdaiincidents2025}, and ``Where in the World is AI?''~\cite{aiglobalmap2025}. 

Additionally, AIES and FaccT communities have significantly contributed to studying instances of real-world harms~\cite{ali2019discrimination,albert2021whole}, and offered recommendations for auditing AI systems~\cite{raji2019actionable}.


\subsection{Research Gap}  
Previous work has focused on classifying AI risks into taxonomies, identifying risks envisioned by researchers, automatically suggesting risks to AI practitioners, and curating harms in the aftermath of real-world incidents.
However, the valuable perspectives of AI developers has been overlooked, limiting our understanding of unique risk profiles of specific AI models and the interactions between their capabilities and common uses. It also means we lack a clear picture of how different experts (e.g., researchers and developers) think about AI risks, and how their concerns match the harms that have already occurred. 

%% file: sections/3_methods.tex
\section{Methods}\label{sec:methods}

To address this research gap, we built and validated the AI Model Risk Catalog based on risks identified by developers, and compared it with risks identified by researchers in the MIT Risk Repository, and with harms reported in the AI Incident Database.

\subsection{Methods for building the AI Model Risk Catalog}\label{sec:methods_buidling_catalog}
We first describe the model cards dataset, and then the methods of using LLMs for extracting risks from this dataset, classifying the extracted risks into taxonomies, and consolidating and validating the classified risks into a catalog.

\subsubsection{Downloading model card snapshots from HuggingFace.}\label{sec:results_downloading_model_cards}
In July 2024, we obtained a snapshot of the model repository from HuggingFace using the HF Hub API\footnote{\url{https://huggingface.co/docs/huggingface_hub/v0.5.1/en/package_reference/hf_api}}. This included 765,973 model repositories, with 461,181 (60\%) model cards. For every model card, we employed regular expressions to identify sections related to risks. Specifically, we searched for mentions of risks, limitations, bias, ethical considerations, out-of-scope uses, misuse, responsibility, and safety. This approach identified 64,116 (14\%) model cards with risk-related sections. Due to the absence of standardized content requirements on HuggingFace, many model cards are incomplete, and numerous risk sections are only slightly modified replicas of one another. Among the $64,116$ model cards, an overwhelming majority of risk sections (96\%) were exact duplicates. We retained $2,672$ model cards with unique risk-related content (selecting the card with highest download count in cases of duplicates) as our \textit{Model Cards up to 2024} dataset.  Furthermore, we also considered a snapshot of all the model cards up to October 2022, released by previous research \cite{liangSystematicAnalysis321112024} to analyse how the risks reported in model cards have evolved over the two years. This snapshot consisted of $74,970$ model repositories and $32,111$ $(42.8\%)$ model cards. We extracted the risk-related sections, resulting in $5,546$ $(17.3\%)$ model cards with any risk-related sections. Among these, 95\% were exact duplicates with 322 model cards having unique risk-related content referred to as \textit{Model Cards up to 2022} dataset.
See Table \ref{tab:data_stats} for details on both data snapshots.

\input{tables/huggingface_dataset}

\subsubsection{Extracting risk mentions.}\label{sec:methods_extracting_risks}
Given the scale of our dataset, manually extracting risks from 2,672 model cards was not feasible. We therefore used a large language model (GPT-4o) to automate the extraction process and validated its accuracy against a manually annotated sample. Large language models have shown strong performance in zero-shot and few-shot annotation tasks~\cite{ziems2024can,strachan2024testing}. We prompted GPT-4o with definitions and examples, asking it to identify whether a section discussed risks and to extract distinct mentions in a verb–object format. To reduce variability and hallucinations, we set the temperature to zero~\cite{peeperkorn2024temperature}. We refined the prompt iteratively until its outputs matched human annotations on a separate 10\% test set, achieving 90\% agreement on a sample of 50 cards. All extracted risks were reviewed by the authors for consistency and accuracy. We removed exact duplicates and further grouped near-duplicates using two methods:  (1) \emph{fuzzy string matching} to detect similar phrasing, and  
(2) \emph{contextual embeddings with cosine similarity}, using the \emph{bge-large-en-v1.5} model~\cite{muennighoffMTEBMassiveText2023}. Pairs with a fuzz score of at least 75 and embedding similarity above 0.85 were treated as duplicates. We retained the longer entry in each pair. These thresholds were calibrated on 5\% of the dataset.

\subsubsection{Taxonomizing risks.}\label{sec:methods_classifying_risks}
After extracting risks from model cards, we classified them using two taxonomies: the DeepMind taxonomy~\cite{weidinger2023sociotechnical} and the MIT Risk Repository taxonomy~\cite{slattery2024ai}. We used GPT-4o to assign each risk to \revision{the most suitable category}. To guide the model, we provided definitions from both taxonomies and asked it to match each risk accordingly. We refined the prompt until its outputs matched manual annotations on an unseen 10\% sample. Because the model’s outputs vary, we classified each risk three times and kept only the categories that appeared at least twice. To evaluate accuracy, we compared the model’s results against a manually annotated sample of 50 risks and found 84\% agreement.



%

\subsubsection{Consolidating risks.}
A manual validation of a sample of extracted risks revealed that the automatic duplicate removal approach was partially effective. To enhance the dataset's credibility, the extracted risks were manually reviewed and refined. For each sub-category of the MIT risk taxonomy, the first author assessed the risks within the category, identified those with significant similarity, and removed duplicates, prioritizing the retention of the risk with the most comprehensive information. These eliminations were then independently verified by the second author. Any disagreements were resolved through discussion. Care was taken to preserve granular details, such as risks specific to particular models or datasets, as well as risks that represented instances of other risks but included additional information. This process resulted in our AI Model Risk Catalog.

\subsubsection{Validating  classification of risks.}
To validate the quality of the risk classifications generated by the LLM, we evaluated its performance on the manually coded MIT risk repository. 
We applied our custom prompt to extract the LLM's classifications of risk mentions in the repository. These predictions were then compared against the repository's manual coding, which served as the ground truth. 
The results showed an accuracy of 83\% and a macro-averaged F1 score of 81\% 
for the seven-class classification task. Most misclassifications occurred in the predicted categories of \emph{malicious actors and misuse} (32 risks), \emph{misinformation} (26 risks), and \emph{AI system safety, failures, and limitations classes} (22 risks). 
Upon closer inspection, we found that these were not necessarily errors. As noted by \citet{weidinger2023sociotechnical} and \citet{slattery2024ai}, these categories are not mutually exclusive, and many risks span multiple categories due to their interconnected nature.
For example, risks ``the demonstrated ability of anonymous actors to accumulate resources online (e.g., Satoshi Nakamoto as an anonymous crypto billionaire)'' and ``this is the risk posed by an ideal system if used for a purpose unintended by its creators'' are classified by LLM under \emph{malicious actors and misuse}, but are found under \emph{AI system safety, failures, and limitations} in the repository. Obviously, the unintended, anonymous, and negative use mentions also justify their inclusion under \emph{malicious uses} category.  Similarly, ``AI-generated or synthesized content can lead to the spread of false information, discrimination and bias, privacy leakage'' could be both risk of \emph{misinformation} (as classified by the LLM) and \emph{discrimination and toxicity} (MIT groundtruth).  

To ensure that the high classification accuracy translates to our data, we conducted a manual review of how all risks are classified in our catalog, paying special attention to the overlapping categories discussed above. The review revealed less than 1\% misclassifications, validating the accuracy and quality of our catalog.

\subsubsection{Thematically analyzing risks.}
 We conducted a thematic analysis \cite{braunThematicAnalysis2012, braunUsingThematicAnalysis2006} of the extracted risks using inductive coding where one author coded the data to comprehend and highlight the characteristics and quality of risk reporting. These codes were then jointly discussed by two authors and resolved for any disagreements.


\subsection{Methods for comparing risk sources}\label{sec:methods_comparing}
%
We downloaded two prominent sources of AI risks, and implemented methods for comparing our catalog to them.

\subsubsection{Sources of researcher-identified risks and real-world harms.}\label{sec:methods_downloading_aiid_mit}
We used two public sources to compare developer-reported risks with those identified by researchers and with real-world harms. First, the MIT Risk Repository\footnote{\url{https://airisk.mit.edu/}} \cite{slattery2024ai} compiles a structured \revision{repository} of 967 risks drawn from 43 AI risk frameworks, categorized across two taxonomies. Second, the AI Incident Database (AIID)\footnote{\url{https://incidentdatabase.ai}} catalogs cases where AI systems have caused harm or failed in practice. Reports are submitted by contributors, reviewed by volunteer editors, and grouped by incident across multiple media sources. As of January 2025, it included 869 incidents based on 4,406 media reports. For brevity, we refer to these sources as the repository (MIT Risk Repository) and the database (AIID) throughout.

\subsubsection{Classifying risks and harms into taxonomies using LLM.}\label{sec:methods_classifying_technique}
For AI Incidents, we followed the method of \citet{velazquezDecodingRealWorldArtificial2024}, classifying each incident description into our two chosen taxonomies. As with model card risks, we provided definitions of the layers and categories from both taxonomies in the prompt. Each incident description was classified three times, and categories that appeared at least twice were kept to ensure consistency. The exact prompts are in Appendix \emph{Prompts} in the pre-print version of this paper. For the MIT Risk Repository, we used the manual labels already provided for the MIT taxonomy and applied the same prompting procedure to map each risk to the DeepMind taxonomy.
This process gave us common coding across all three sources (model cards, MIT repository, and AIID), letting us compare how risks envisioned by developers and researchers aligned or diverged from those shown as harms in the news.

\subsubsection{Comparing risk sources.}\label{sec:methods_comparing_technique}
To compare the prevalence of risk categories across the three sources, we calculated confidence intervals for the difference of proportions between each category using the Miettinen-Nurminen asymptotic score method~\cite[p. 250]{fagerland_recommended_2015}. This analysis employed the \texttt{diffscoreci()} function from the \texttt{PropCI} package in R~\cite{scherer_propcis_2018}. Following prior research~\cite{salehzadehniksiratChangesResearchEthics2023}, we chose this method over Z-tests, as the boundary probabilities (close to 0 or 1) in some categories render Z-tests and their confidence intervals unreliable~\cite[p. 164]{agresti_score_2011}.

%% file: tables/huggingface_dataset.tex
\begin{table*}[t!]
\centering
\footnotesize
\caption{Statistics of two HuggingFace model card snapshots that we used for creating two versions of the AI Model Risk Catalog (2022 and 2024). Out of all the model cards, very few have a completed risk section, and an even smaller number have unique risk sections (i.e., sections that are not copied from other cards). 
}
\begin{tabular}{@{}lccc cc@{}}
\toprule
& \multicolumn{3}{c}{Stats for all model cards} & \multicolumn{2}{c}{Stats for the model cards with unique risk sections} \\ 
\cmidrule(lr){2-4} \cmidrule(lr){5-6}
Dataset & Total cards & With risk sections & With unique risk sections & Average \# downloads & Characters in risk sections \\ \midrule
\emph{Model cards up to 2024} & 461,181 & 64,116 & 2,672 & 118,508.41 & 759.69 \\ 
\emph{Model cards up to 2022} & 32,111 & 5,546 & 322 & 364,639.04 & 882.90 \\ 
\bottomrule
\end{tabular}
\label{tab:data_stats}
\end{table*}

%% file: sections/4_results.tex
\section{AI Model Risk Catalog}\label{sec:results}

We built the catalog using two snapshots of Hugging Face model cards dated through 2022 and 2024. We then examined how many risks developers report, how they report them, what kinds of risks they describe, and how their reporting has changed over time.

\subsubsection{How many risks developers report?}\label{sec:results_caalog_stats}
In the snapshot of model cards up to 2024, of the 64,116 cards with risk-related sections, 54,448 (approximately 85\%) lacked substantive risk information. Most of these either simply retained the default template requesting risk details ($52,983$), or referred to another related card for risk content. The remaining $9,668$ cards contained a total of $37,401$ standardized risk mentions, including duplicates.

We removed duplicates by keeping the most downloaded model cards, resulting in a final dataset of $2,672$ cards with unique risk content. From these, our extraction method produced $3,588$ distinct, standardized risk mentions. 
%
The consolidation of risks (including de-duplication and streamlining) reduced the preliminary set of $3,588$ risks to the final set of $2,863$ risks ($\sim 20\%$ reduction). 

In the snapshot of model cards up to 2022, of the 4,546 cards with risk-related sections, 367 lacked risk information. The remaining $4,179$ cards contained a total of $9,645$ standardized risk mentions, including duplicates.
Removing duplicates resulted in 322 cards with 474 standardized unique risk mentions. 

\subsubsection{How developers report risks?}\label{sec:how}
\noindent The \emph{thematic analysis} of all the extracted risks revealed five themes related to \textit{risk reporting practices of developers}:
    \begin{enumerate}[label=\arabic{enumi}., leftmargin=15pt, topsep=3pt]
        \item \emph{Repetition and redundancy}: Although expressed in many different phrasings, the risk mentions tend to cover similar ground, indicating pervasive concerns among AI developers. Those include bias and fairness, output quality and accuracy, safety and harmful content, privacy and security, operational and technical limitations, as well as some ethical and societal implications.
        \item \emph{Ambiguity}: In line with previous research \cite{bhatAspirationsPracticeML2023, crisan2022interactive}, many risk mentions are ambiguous, lack specificity, or generally addresses a large type of models. Some examples are {``is not immune from issues that plague modern large language models''}, {``generates confusion''}, {``makes mistakes''}, and {``increases risk to users if used irresponsibly''}.
        \item \emph{Granularity levels and specificity}: In addition to the risk mentions such as those above that are very general, others are very capability-specific. For instance, for a general risk saying that the ``model underperforms on out-of-distribution data,'' specific versions could be saying that it does so with particular languages, dialects, input types (e.g., ``jpeg artifacts''), or for certain classification labels (e.g., emotion ``fear''). 
        \item \emph{Interdisciplinary concerns}: The risks span across multiple dimensions of AI use, from code generation and image synthesis to language understanding and ethical decision-making, highlighting that developers understand that these challenges are not isolated but rather systemic and interconnected.
        \item \emph{Warnings for deployment}: There is an overall emphasis on caution, with many entries explicitly advising against using the models in critical or unvetted applications without robust human oversight, rigorous testing, and additional safety mechanisms. Examples include {``should not be used as a substitute for professional legal advice''}, {``should not be solely relied upon for real-time critical medical decisions.''}
    \end{enumerate}

\subsubsection{What types of risks developers report?}\label{sec:cat}
To understand the types of model risks developers imagine, and align with prior work~\cite{slattery2024ai,weidinger2021ethical}, we opted to classify the risks using established taxonomies. The DeepMind taxonomy~\cite{weidinger2023sociotechnical}---currently the most widely cited in the literature, as noted by \citet{slattery2024ai}---served as a natural starting point. Additionally, we found that an augmented version of this taxonomy, developed as part of the MIT Risk Repository efforts~\cite{slattery2024ai}, introduced an extra category (\emph{AI system safety, failures, and limitations}) applicable to many of the risks in our catalog. Moreover, this augmented taxonomy facilitates a direct comparison between our findings and those in the MIT Risk Repository, currently the largest AI risk repository. For these reasons, we chose the MIT risk taxonomy as the second one to classify our risks with.

Appendix Figure~\ref{fig:mit_classification} show how the cataloged risks map onto the MIT risk taxonomy. The most common category is \emph{discrimination and toxicity} (44\%), with 24\% of all risks tied to \emph{unfair discrimination and misrepresentation}, followed by \emph{exposure to toxic content} (8\%) and \emph{unequal performance across groups} (8\%). These risks often refer to biased training data and skewed model outputs on fairness benchmarks~\cite{le2022survey}. The next most frequent category is \emph{AI system safety, failures, and limitations} (37\%), with most of these risks falling under \emph{lack of capability or robustness} (35\%). They include reports of model bugs, inefficiencies, or errors in output quality, reflecting a primarily technical view among developers. The \emph{misinformation} category focuses on hallucinated or inaccurate information. \emph{Privacy and security} risks relate to code vulnerabilities, unauthorized sharing of copyrighted material, and memorization of training data. Though less common, \emph{malicious actors and misuse} includes concerns about models enabling illegal activity or violating human rights. The \emph{overreliance and human agency issues} category addresses AI use in sensitive domains like healthcare and the risks of anthropomorphizing models, though it accounts for less than 2\% of all risks. Similarly, the \emph{socioeconomic and environmental harms} category—also under 2\%—covers the use of computational resources, legal compliance, and broader societal impacts.

These risk patterns reflect the dominance of language models on Hugging Face (62\%), which are widely known for raising concerns about discrimination and misinformation~\cite{10.1145/3442188.3445922, ousidhoum-etal-2021-probing}. However, our analysis by input and output modality, shown in Appendix Table~\ref{tab:risk_mentions}, reveals more nuances and details. Non-text inputs are more often linked to \revision{\emph{socioeconomic and environmental harms} (22\% of such risks), and \emph{privacy and security} (20\%). In total, 9\% of all risks involve non-text inputs, and 17\% involve non-text outputs}. These risks span thousands of models and reflect a known gap in safety evaluation for non-text modalities~\cite{weidinger2023sociotechnical}. The risks identified in our catalog could serve as an inspiration or starting point while addressing this gap, and could help guide future work in this area.

We also find that different risk categories become more prominent with \emph{multimodal} models. For models with multimodal input, compared to other risk categories, developers report more \emph{malicious use}  \revision{($14$\% of that category)}, as well as \emph{privacy and security} risks ($13$\%).
These findings show how risk profiles shift when looking beyond text-based models.

As shown in {Appendix} Figure \ref{fig:deepmind_classification}, the distribution of risks across DeepMind's taxonomy reflects similar main trends described when using the MIT taxonomy. However, because the DeepMind taxonomy lacks a category for model-specific risks—\emph{AI system safety, failures, and limitations}—some risks in our catalog could not be clearly classified and were instead all grouped under \emph{representation and toxicity harms}, which in this case accounts for over 62\%. For more clarity, and easier comparison with the MIT Risk Repository, we report the rest of our findings in main text using the MIT taxonomy.

\subsubsection{How developers’ risk reporting has evolved?}\label{sec:compariosn_2022_2024}
\begin{figure*}[t]
    \centering
    \includegraphics[width=0.99\textwidth]{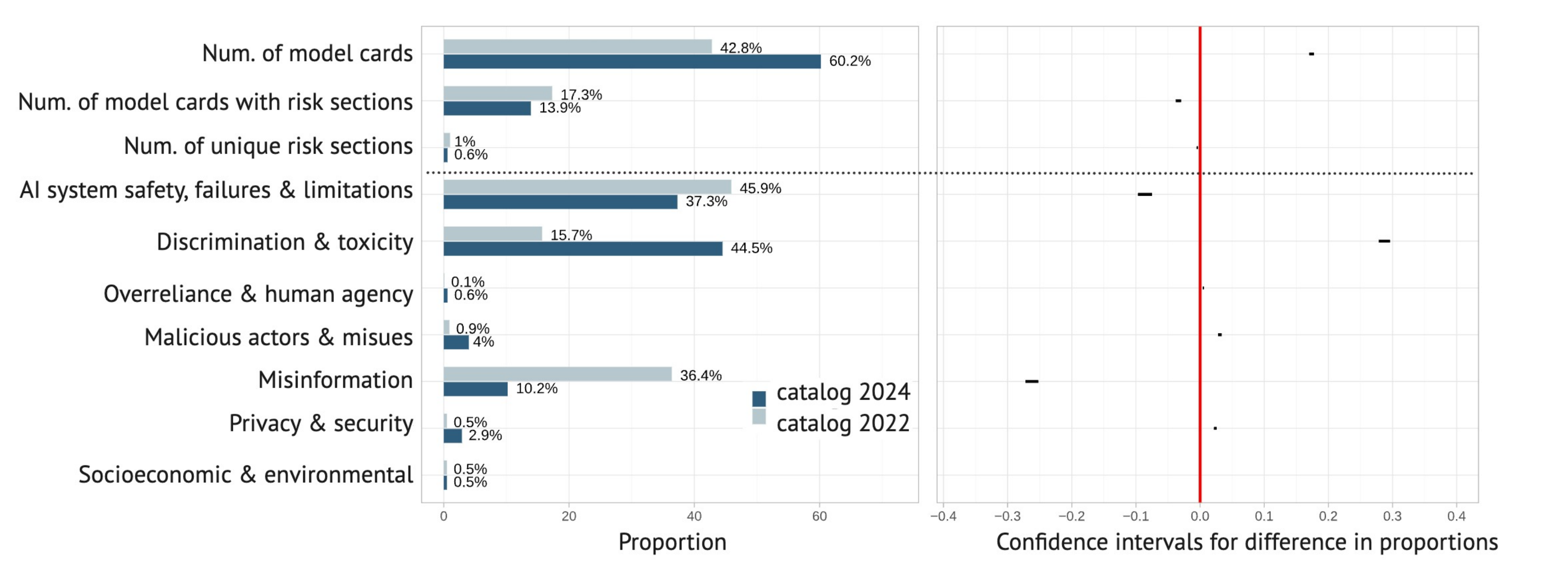}

    \caption{Comparing AI Model Risk Catalogs from 2022 and 2024. \emph{Left:} The percentages of risks in the catalog that belong to each category. \emph{Right:} 95\% confidence intervals for the pairwise differences between each pairs of percentages. A greater distance of the interval from the vertical red line indicates a larger difference, while a narrower interval reflects higher certainty.
    The top section (above the dashed line) shows statistics from the model cards snapshot, while the bottom section presents the seven risk categories. Although the number of models with model cards has grown significantly, the proportion with risk sections has dropped even further (from 17\% to 14\%). The \emph{AI system safety} risks were the most prevalent in 2022, overtaken by \emph{discrimination and toxicity} in 2024. There is a threefold decrease in the proportion of \emph{misinformation}, and fourfold increase in the \emph{malicious uses} risks between the two years. The remaining categories have remained mostly stable.} \label{fig:results_comparing_2022_2024}
\end{figure*}
Previous research analyzing the 2022 snapshot of model cards \cite{liangSystematicAnalysis321112024} reported that developers often struggle to complete the risk sections: less than $17\%$ of cards had any risk content reported. To understand if risk reporting has changed since then, in Figure~\ref{fig:results_comparing_2022_2024}, we compare the 2022 and 2024 model card snapshots, and our two versions of the AI Model Risk Catalog based on those snapshots. 
While there is a significant increase in the total number of model cards over this period, the percentage of cards with completed risk sections has \emph{decreased} to $14\%$, as has the percentage with unique risk sections ($<1\%$). We speculate this decrease happened because newer models are often derived from or fine-tuned versions of foundational models like GPT, and may end up reusing their original risk sections. Additionally, the average length of the risk sections has also declined (see Table~\ref{tab:data_stats}).
Our findings not only confirm those of earlier studies that highlight developers' struggles with risk communication ~\cite{liangSystematicAnalysis321112024,bhatAspirationsPracticeML2023,crisanInteractiveModelCards2022}, but they raise a critical concern that the situation may have even worsened over time.

In terms of the types of risks developers envision, Results in Figure~\ref{fig:results_comparing_2022_2024} indicate a relative stability overall. The notable difference is a swap in the two most frequent categories: in 2022, \emph{AI system safety, failures and limitations} were most prevalent, followed by \emph{discrimination and toxicity}; in 2024, this order has reversed. Also, the number of \emph{misinformation} risks has significantly reduced in 2024 (threefold decrease), whereas those about \emph{malicious actors and misuse} have increased (fourfold increase, though still accounting for only 4\% of the total risks), as well as those about \emph{privacy and security} (sixfold increase, accounting to 3\% of the total risks now). This is likely related to the increase in the multimodal models (see ~\citep{10.1093/nsr/nwae403}, and our Appendix Figure \ref{fig:comparing_2022_2024_modalities}), which are linked with more of such risks, as we discussed in the previous section. The prevalence of risks in other categories has not changed significantly.

%% file: sections/5_results.tex
\section{Comparison of risks envisioned by AI developers and researchers with harms recorded in the news}
\label{sec:results_v2}
\begin{figure*}[t]
    \centering
    \includegraphics[width=0.99\textwidth]{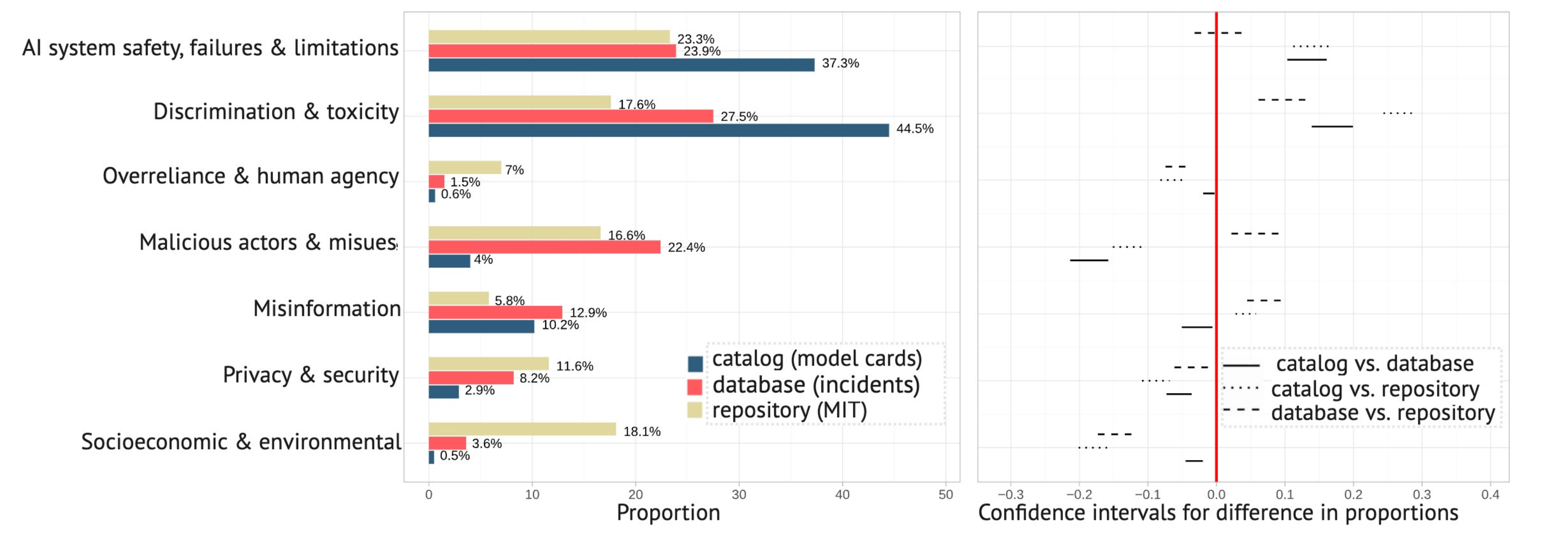}
    \caption{Comparison of three risk sources. We compare our catalog (developer-envisioned risks in model cards), repository (researcher-envisioned risks in the MIT Repository), and the database (harms recorded in the AI Incidents). \emph{Left:} The percentages of risks in the catalog / database / repository that belong to each category. \emph{Right:} 95\% confidence intervals for the differences between each pair of percentages. A greater distance of the interval from the vertical red line indicates a larger difference, while a narrower interval reflects higher certainty.}
    \label{fig:three_comparisons}
\end{figure*}
%


Having built the AI Model Risk Catalog, we compared risks reported by developers, those identified by researchers, and harms recorded in real-world incidents. Figure~\ref{fig:three_comparisons} shows the results, we highlight three key takeaways next. Note that, to capture the full range and frequency of developer-reported risks, we analyzed all 37,401 mentions, including duplicates,  from 64,116 model cards with risk sections, rather than limiting our analysis to the 2,863 unique entries in the catalog. This approach aligns with the MIT Risk Repository and the AI Incident Database, where the same risks or harms may appear in multiple research papers or news reports.

%

\vspace{1em}
\noindent
\textbf{Takeaway 1:} \emph{Developers imagine more of the risks tied to AI capabilities (37\%) and bias (over 44\%) than their share in real-world recorded harms (24\% and 27\%). Still, these risks account for over half of the harms, which supports the relevance of developers' concerns.}

When developers focus on risks tied to model capabilities (\emph{AI system safety, failures and limitations}, 37\%), it reflects their hands-on role: they build, test, and evaluate model performance. They tend to report concrete technical issues (e.g., ``misclassification due to tokenization,'' ``context window saturation,'' or ``texts longer than 128 tokens'') along with problems in how models process information, the quality of outputs, and behavior in specific settings. Unlike risks in the research repository or the incident database, the developer-reported catalog highlights fine-grained model behaviors in controlled environments. 
As shown in Appendix Figure~\ref{fig:mit_classification}, most of these developer-imagined risks fall under the subcategory of \emph{lack of capability or robustness}, which also accounts for most real-world harms. The other two subcategories—\emph{lack of transparency or interpretability} and \emph{AI pursuing its own goals}—each account for less than 1\% of developer-reported risks and real-world harms. Still, this points to a potential gap: transparency and interpretability are key concerns in responsible AI~\cite{BARREDOARRIETA202082,samek2017explainable}, especially when it comes to the adoption in high-stakes domains such as healthcare~\cite{amann2020explainability}. Notably, researchers give these risks slightly more attention (2.6\%), as shown in the MIT Risk Repository distribution in Appendix Figure~\ref{fig:three_comparisons}.

Bias-related risks (\emph{discrimination and toxicity}) make up the majority (over 44\%) of developer-reported risks and 27\% of real-world harms. This shows that developers are not only concerned with model capability, but also with value alignment, especially around fairness and the risk of producing discriminatory or toxic content. As discussed earlier (see Figure~\ref{fig:results_comparing_2022_2024}), these concerns have more than doubled since 2022, showing that developers are paying increasing attention to discrimination from their models. These risks are often tied to specific model types and modalities, such as speech-to-text or audio-visual models. Developers also anticipate use-specific risks: bias in models used in healthcare, hiring, or other sensitive areas is commonly mentioned. Some catalog entries are highly specific; for example, inherited bias from a certain dataset or ``biases toward anime female characters.'' Among the three bias-related subcategories (Appendix Figure~\ref{fig:mit_classification}), most developer-reported risks fall under \emph{unfair discrimination and misrepresentation} (24\%), followed by \emph{exposure to toxic content} (12\%) and \emph{unequal performance across groups} (8\%). 

Since 24\% of recorded harms relate to \emph{AI system safety} and 27\% to \emph{discrimination and toxicity}, the concerns raised by developers are valid and timely. One example of the first type of harm involves a warehouse robot that ruptured a can of bear spray and injured workers (incident 2). An example of the second type is when Google Image search showed a Barbie doll as the first female ``CEO''—after 11 rows of male CEOs (incident 18).

\vspace{1em}
\noindent \textbf{Takeaway 2:} \emph{Compared to recorded real-world harms, AI researchers imagine more of the risks from three categories: socioeconomic and environmental harms (18\%), overreliance and human agency (7\%), and privacy and security (12\%). These risks appear nearly three times more often in research than in incident data, and over ten times more than in developer-envisioned risks.}

Only 4\% of real-world harms fall under \emph{socioeconomic and environmental harms}, compared to 18\% of researcher-imagined risks. We see this strong focus by researchers not as misplaced but as forward-looking. As noted by \citet{velazquezDecodingRealWorldArtificial2024}, systemic harms often take longer to materialize and are less likely to be reported in the news—not because they are less serious, but because their effects are slower and harder to trace. Researchers give roughly equal weight (Appendix Figure~\ref{fig:mit_classification}) to subcategories such as \emph{power centralization and unfair distribution of benefits} (4\%), \emph{increased inequality and decline in employment quality} (3\%), \emph{economic and cultural devaluation of human effort} (2\%), and \emph{environmental harm} (3\%). By contrast, developers imagine all these risks rarely—between 0.1\% and 0.3\%.

For \emph{overreliance and human agency}, researchers envision 7\% of their reported risks in this category, while it makes up only 2\% of real-world harms, and 0.6\% of developer-imagined risks. Researchers focus on both \emph{overreliance and unsafe use} (4\%) and \emph{loss of human agency and autonomy} (3\%). These risks may signal early signs of systemic issues—for example, overreliance leading to new types of mistakes at work, or loss of agency affecting worker well-being. Hence, once more, we see researchers' attention to these concerns as justified and forward-thinking.

The higher share of \emph{privacy and security} risks in research is also notable (12\% compared to 8\% of real-world harms and only 3\% of developer-imagined risks). While researchers envision equal proportions (6\%) of \emph{AI system security vulnerabilities and attacks} and \emph{compromise of privacy by leaking sensitive information}, it is the latter category that results in more incidents than the former (7\% \emph{vs.} 1\%).

\vspace{1em}
\noindent \textbf{Takeaway 3:} 
\emph{Both researchers and developers tend to overlook risks tied to malicious use and misinformation. Harms of this kind linked to human interaction  and social engineering are significantly more common in incidents than in expert reports, indicating blind spots among the experts.}

\emph{Malicious uses} of AI are widespread, accounting for over 22\% of recorded harms in the incident database, yet they receive less attention from researchers (17\%) and especially developers (4\%). The most common subcategories are \emph{fraud, scams, and targeted manipulation} (15\%) and \emph{disinformation, surveillance, and large-scale influence} (7\%).
One example of fraud and manipulation is the use of deepfake videos to scam Canadian immigrants out of thousands of dollars (incident 876). A case of political disinformation involves a deepfake that falsely showed U.S. Congressman Rob Wittman endorsing military support for Taiwan’s Democratic Progressive Party ahead of the 2024 election (incident 876).
These harms are often tied to specific model features~\cite{charfeddine2024chatgpt, golda2024privacy} and their data sources~\cite{liu2024generative}. Hence, the low number of developer-identified risks in this area shows a clear blind spot. One example from the catalog is the risk ``outputs realistic faces,'' linked to a text-to-video model. This risk could cause the above mentioned harms. But this description is broad and vague—it could apply to many harms and is not helpful for mitigation.
A more detailed researcher-identified risk notes that ``GenAI can produce images of people that look very real, as if they could be seen on platforms like Facebook, Twitter, or Tinder. Although these individuals do not exist in reality, these synthetic identities are already being used in malicious activities.'' However, as with the developer risk, there are many possible misuses of GenAI that this risk could enable, which likely explains the gap between the proportions of actual harms and expert-envisioned risks in this category.

Although there is a growing body of research on \emph{misinformation}~\cite{chen2024combating, liu2024preventing, garimella2017image}, the actual harms seen in the world (13\%) suggest that these risks are more serious and more frequent than the developers (10\%), and especially researchers (6\%) envision.
Harms of both of these types can be linked to human interaction~\cite{velazquezDecodingRealWorldArtificial2024}, and social engineering \cite{Wang2021,Schmitt2024} highlighting at a gap in expert focus.

%% file: sections/6_discussion.tex
\section{Discussion}\label{sec:discussion}
%

By curating risks from nearly half a million model cards into an \emph{AI Model Risk Catalog}, we help clarify how developers see AI risks. We also compare the focus of developers and researchers with real-world harms reported in the news, revealing where they overlap and differ, and where important gaps remain. 
Below, we outline the implications of our work for bridging the divide between these groups and other stakeholders, such as journalists, policy-makers, auditors, and the wider public.

\subsection{Theoretical implications}
\noindent
\textbf{Definition and format of AI risk.} 
Our comparison of risks from two leading sources (the repository and the incident database) with those in our new catalog shows there is no standard way to report AI risks. Some risks are described in long, narrative statements that cover not just failures or incidents but also broader consequences—such as ethical dilemmas, moral challenges, or legal issues (mainly in the repository). Others are brief and sometimes vague statements about what could go wrong (often in the catalog). Real-world harms, by contrast, are tied to specific uses, domains, and subjects harmed (as in the database). 

This points to a clear need for the responsible AI community to adopt a structured standard for defining and communicating risks. Drawing on insights from all three sources, we argue that a robust risk format should at minimum specify the situation and context in which harm might occur. Without this, a single expert-defined risk can lead to many different harms of varying severity and likelihood (as we have discussed in \emph{Results}). Such a standard would enable interoperability across sources and help all stakeholders—developers, policy-makers, auditors, and users—understand and act on risk information. A recent step in this direction is the Risk Card format~\cite{derczynski2023assessing} for language model risks. Still, further simplification and user studies are needed to ensure risk communication meets the needs of different audiences.


\mbox{ } \\
\noindent
\textbf{Unified source of AI risks and harms.} 
Our results show that these three resources complement each other: the repository and database provide depth of insight and examples of broader, systemic harms, while the catalog together with risk links to specific models and datasets, can serve as a granular reference tool for the systematic auditing of AI system capabilities~\cite{uuk2024effective}.
Together, these resources create a shared knowledge base that can support collaboration across a wide range of stakeholders.

However, while we cover two key expert groups and the media, important perspectives are still missing. These include the views of end users and affected communities, civil society organizations and NGOs working in digital rights, consumer protection, and human rights, as well as legal experts, ethicists, and sectoral specialists from fields like healthcare, finance, and energy.
Involving these groups is crucial for building a fuller, more inclusive understanding of AI risks and ensuring that risk reporting and mitigation efforts meet the needs of all those affected.

\subsection{Practical implications}

\noindent
\textbf{Researchers and scholars.} 
Our findings show that researchers have driven the discussion on the broader societal, environmental, and governance impacts of AI. However, they should place greater focus on risks linked to malicious use, misinformation, discrimination, i.e., especially those rooted in human interaction and social engineering~\cite{Schmitt2024}. While there is a strong body of work on misinformation~\cite{chen2024combating, liu2024preventing} and bias~\cite{o2024gender, whittaker2019disability, dai2024bias, yu2024large}, these issues remain urgent due to the frequency and severity of real-world harms, as shown by incident data. Social engineering, in particular, exploits human psychology to deceive individuals and groups, often leading to significant harm~\cite{Wang2021}.

To address these challenges, researchers should develop new frameworks and practical solutions for AI risks rooted in human behavior and social dynamics. This requires closer collaboration with the human-computer interaction (HCI) community~\cite{10.1145/3544549.3583178}, and experts in social and behavioral sciences~\cite{WASHO2021100126}. Such partnerships can help anticipate new threats of these types, and design interventions grounded in real user experience.

\mbox{ } \\
\noindent
\textbf{Developers.}
We found that model developers, much like researchers, often overlook risks related to human interaction, social engineering, malicious use, and misinformation. However, developers also show a distinct blind spot for privacy and security risks. This lack of attention is especially puzzling since privacy, security, and malicious use are directly tied to specific model designs, underlying data, and the amplified vulnerabilities introduced by LLMs~\cite{charfeddine2024chatgpt,liu2024generative,bullwinkel2025lessons}.

Bridging these gaps will require developers to systematically consider a broader set of risks, both technical and human-centered. 
First, using risk taxonomies as assessment guides or ``cheatsheets'' can support more complete and systematic developer risk assessments. 
Second, we recommend drawing lessons from established cybersecurity and privacy practices and approaching AI systems as software that can be compromised. Security information sharing and the adoption of security-first and privacy-first approaches—as advocated by the NIST AI management framework~\cite{ai2024artificial,ai2023artificial}—are key strategies~\cite{uuk2024effective}. Third, developers can also benefit from tools like RiskRAG~\cite{rao2025riskrag}, which guide risk thinking in relation to real-world AI uses. Future automatic tools for envisioning risks should ground suggestions in evidence from our catalog, repositories like the MIT AI Risk Repository, incident databases, and structured resources such as BenchmarkCards~\cite{sokol2024benchmarkcards}, rather than relying solely on LLMs.

\mbox{ } \\
\noindent
\textbf{Media professionals.}
Our findings can help media professionals reflect on how they cover AI risks and harms. Rather than focusing only on high-profile or sensational incidents, media outlets can broaden their reporting to include a wider variety of risks and affected groups. By highlighting both common and overlooked harms, media professionals can play a key role in informing researchers and the public, shaping balanced discourse (e.g., by following UNESCO’s guide to best practices in AI reporting \cite{unesco}), and holding developers and policymakers accountable.



\mbox{ } \\
\noindent
\textbf{Policy makers.} 
%
Policymakers can benefit from these risk sources by drawing on risks identified by each of developers, researchers, and in real-world incidents when shaping legal frameworks~\cite{golpayegani2024ai}. Consolidating diverse risks in this way helps pinpoint areas needing urgent attention and can guide resource allocation. These structured resources support the development of targeted regulations, compliance tools~\cite{bogucka2024ai,golpayegani2024ai}, and effective mitigation strategies~\cite{uuk2024effective}.

Unlike less detailed sources, our catalog links risks to specific models, making it more useful for compliance checks and integration into compliance platforms. Our catalog also tracks how often each risk is reported, highlighting the most common concerns across different model types and modalities.  Notably, as multimodal models become more widespread, they introduce new safety challenges~\cite{weidinger2023sociotechnical,hameleers2020picture}, and our results show these models are rapidly increasing in number among the open-source ones on Huggingface platform—an area that requires special attention from policymakers.
Even though many HuggingFace models lack risk documentation, sharing model-specific risks in a public repository like ours can aid model selection, mandatory impact assessments~\cite{bogucka2024co,bogucka2024ai}, and inform product choices for end users~\cite{bogucka2024atlas}.
Lastly, by adopting and encouraging open risk reporting standards, policymakers can further promote transparency and public trust in AI.

\mbox{ } \\
\noindent
\textbf{Public.}
Human-interaction and social engineering risks deserve special attention, and preventing these harms requires greater public awareness. The risk sources such as our catalog can help the public understand how AI might affect them~\cite{bogucka2024atlas}, support informed choices about adopting technology, and encourage accountability among developers, organizations, and policymakers. Public engagement in risk reporting, e.g., through participatory auditing, and discussion can strengthen accountability, and improve AI integration into society.

\subsection{Limitations}

\noindent
\textbf{Limitations and biases of the risk sources.}
We treated the MIT Risk Repository as a source of risks identified by researchers, the AI Model Risk Catalog as reflecting risks described by developers, and the AI Incident Database as evidence of real-world harms. These categories are not clear-cut: many developers are also researchers, and their roles often overlap. The repository includes only a subset of academic work; most model cards do not mention risks, limiting developers’ perspectives; and the incident database may reflect media coverage biases~\cite{shaikh2024recognize,brennen2018industry}. While Hugging Face is one of the most comprehensive public sources of model documentation, our catalog includes only its model cards. It excludes risks described in other platforms, such as GitHub repositories or internal company records~\cite{fang2020introducing}. Finally, the catalog reflects a snapshot from July 2024. It does not capture model cards added or revised since then. Our findings should be read in light of ongoing updates to Hugging Face.

\mbox{ } \\
\noindent
\textbf{LLM shortcomings.} 
Generated risks may involve inaccuracies from LLM hallucinations~\cite{mittelstadt2023protect}. To mitigate this, we extracted and minimally standardized risk sentences to limit misrepresentation.  Potential biases in LLM-based classifications~\cite{luccioni2024stable} were addressed by repeating classifications thrice and applying majority agreement, with manual validation ensuring reliability.

%% file: sections/statements.tex

\section{Adverse Impact Statement}
The proposed risk catalog has the potential for dual-use. Any documentation reporting failure modes and risks could be leveraged by malicious actors to scale harmful or dangerous applications of AI systems. However, the risks included in the catalog are derived from publicly available sources, i.e., model cards, and do not introduce new or undisclosed vulnerabilities. By consolidating and categorizing these risks, our aim is to support responsible AI development and risk mitigation, while minimizing the potential for misuse through careful curation and transparency.

\section{Ethical Considerations Statement}
This work adheres to ethical research principles and community norms, ensuring compliance with applicable laws and professional ethical codes. The data used in our study were sourced entirely from publicly available model cards on HuggingFace and established repositories like the MIT Risk Repository and AI Incident Database. HuggingFace allows the use, display, publication, reproduction, distribution, and creation of derivative works according to their terms of service. We did not collect or utilize sensitive user data, nor did we interact with users or deploy systems during the research.

To mitigate ethical concerns, we took several precautions. First, we acknowledge the potential dual-use risks of documenting AI failure modes and risks, which could be exploited for harmful purposes. However, all risks in our catalog were derived from publicly accessible data, and no new or undisclosed vulnerabilities were introduced. Second, we addressed biases and inaccuracies inherent to LLM outputs through repeated classification runs, majority voting, and manual validation. Finally, we made efforts to ensure the curated catalog contributes to responsible AI development by consolidating risks in a way that supports transparency and accountability.

\newpage

%% file: appendix.tex
\section{Appendix}



\lstdefinestyle{mystyle}{
    basicstyle=\ttfamily\tiny,
    frame=none,
    breaklines=true,
    stepnumber=1,
    numbersep=5pt,
    keywordstyle=\color{blue},
    commentstyle=\color{green},
    stringstyle=\color{red},
    showstringspaces=false
}

\lstset{style=mystyle}

\subsection{(A) AI Model Risks Catalog} \label{appn:ai_model_risk_catalog}
\input{tables/top_risks}

\begin{figure*}[t]
    \centering
    \includegraphics[width=0.97\textwidth]{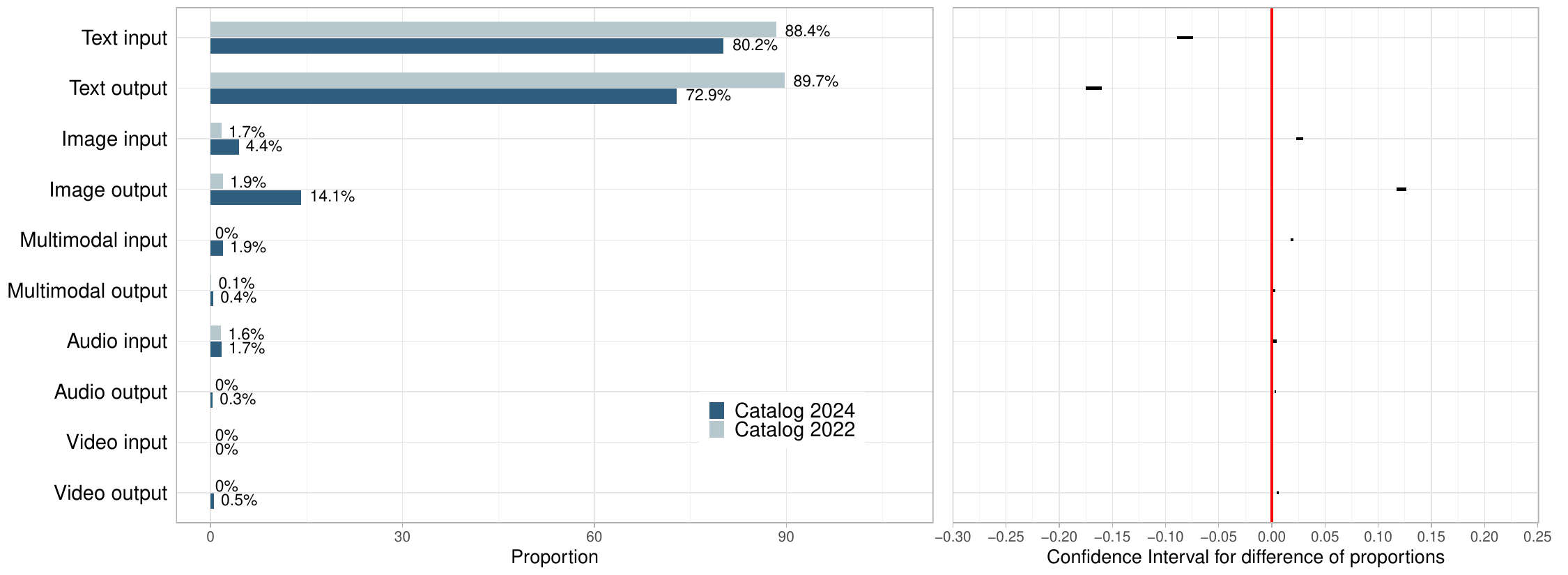}

    \caption{Comparing the AI Model Risk Catalogs from 2022 and 2024 by model input and output types. \emph{Left:} The percentages of risks in the catalog that belong to each category. \emph{Right:} 95\% confidence intervals for the pairwise differences between each pairs of percentages. A greater distance of the interval from the vertical red line indicates a larger difference, while a narrower interval reflects higher certainty.
    Both text input and output have declined since 2022, but still make up 70–80\% of modalities. This reflects a more than threefold rise in image inputs and outputs. Multimodal input has also grown quickly, from almost zero to 2\%. These results show a clear trend toward more non-textual modalities~\cite{10.1093/nsr/nwae403}.} \label{fig:comparing_2022_2024_modalities}
\end{figure*}

\clearpage
\newpage 
\subsection{(B) Comparing risks across sources} \label{appn:comparison_risk_sources}
\begin{figure*}[ht!]
    \begin{flushright}
    \includegraphics[width=0.84\textwidth]{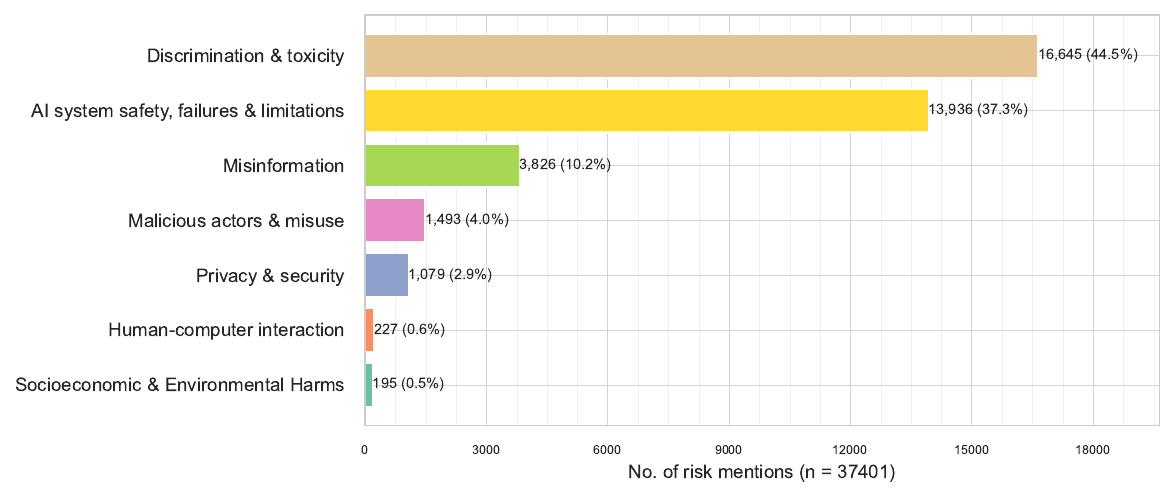}
    
    \includegraphics[width=0.96\textwidth]{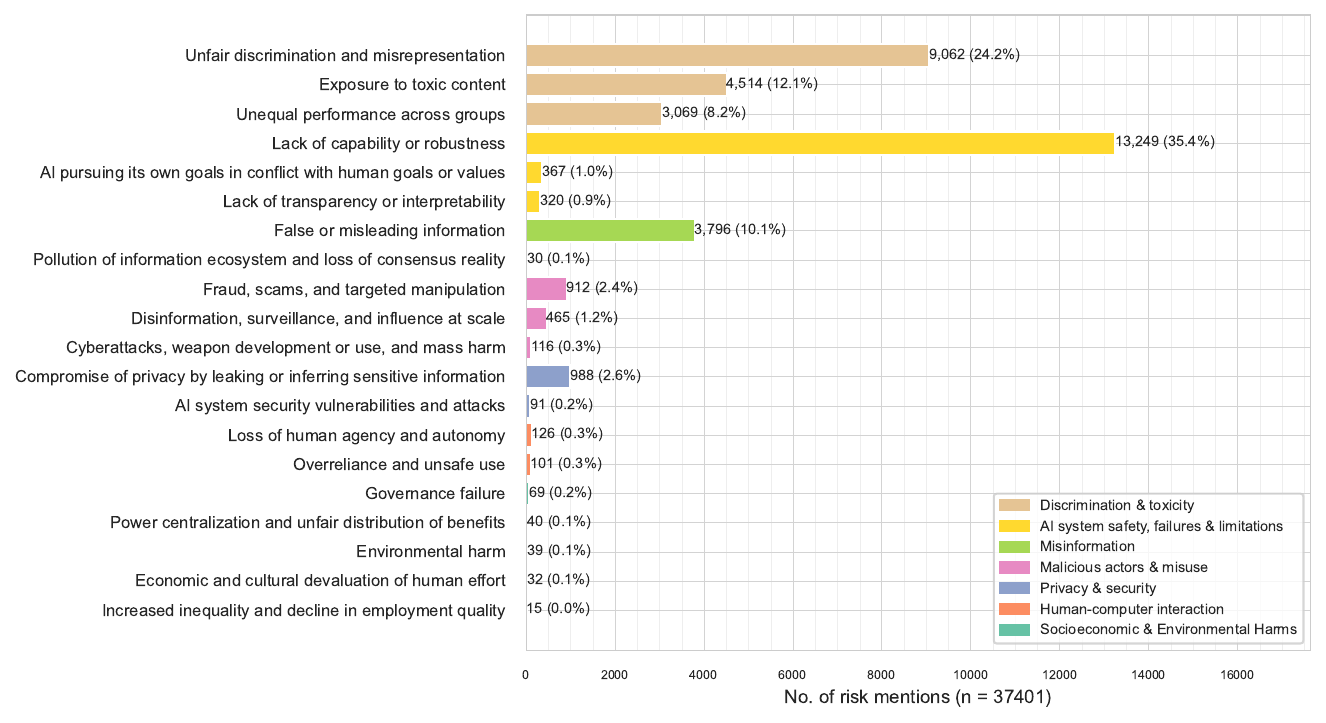}\hfill
    \caption{Proportions of risks in the \emph{AI Model Risk Catalog} falling into different \emph{MIT taxonomy}~\cite{slattery2024ai} categories (top) and subcategories (bottom).} 
    \label{fig:mit_classification}
    \end{flushright}
\end{figure*}
\begin{figure*}[ht!]
    \begin{flushright}
    \includegraphics[width=0.84\textwidth]{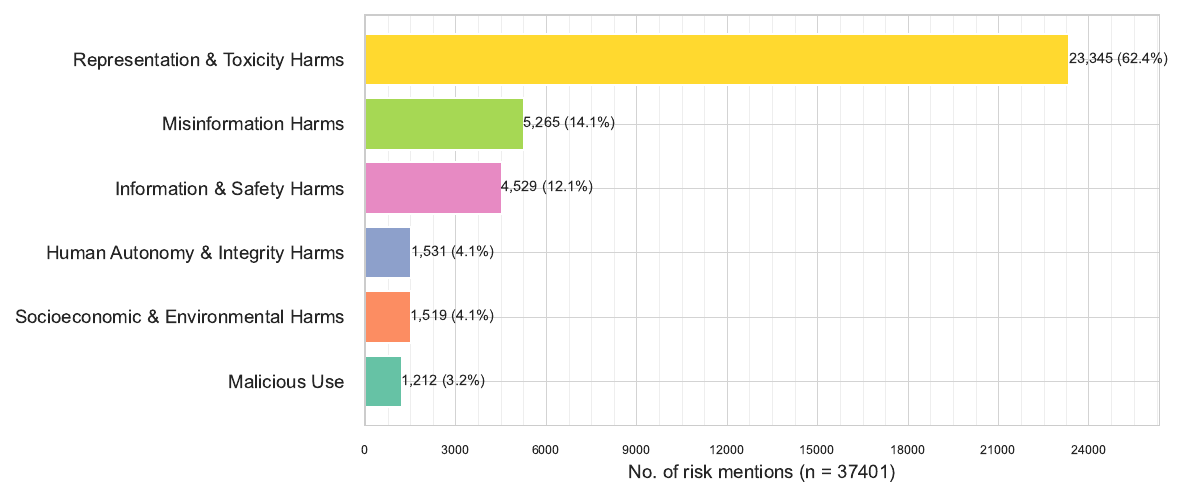}
    
    \includegraphics[width=0.99\textwidth]{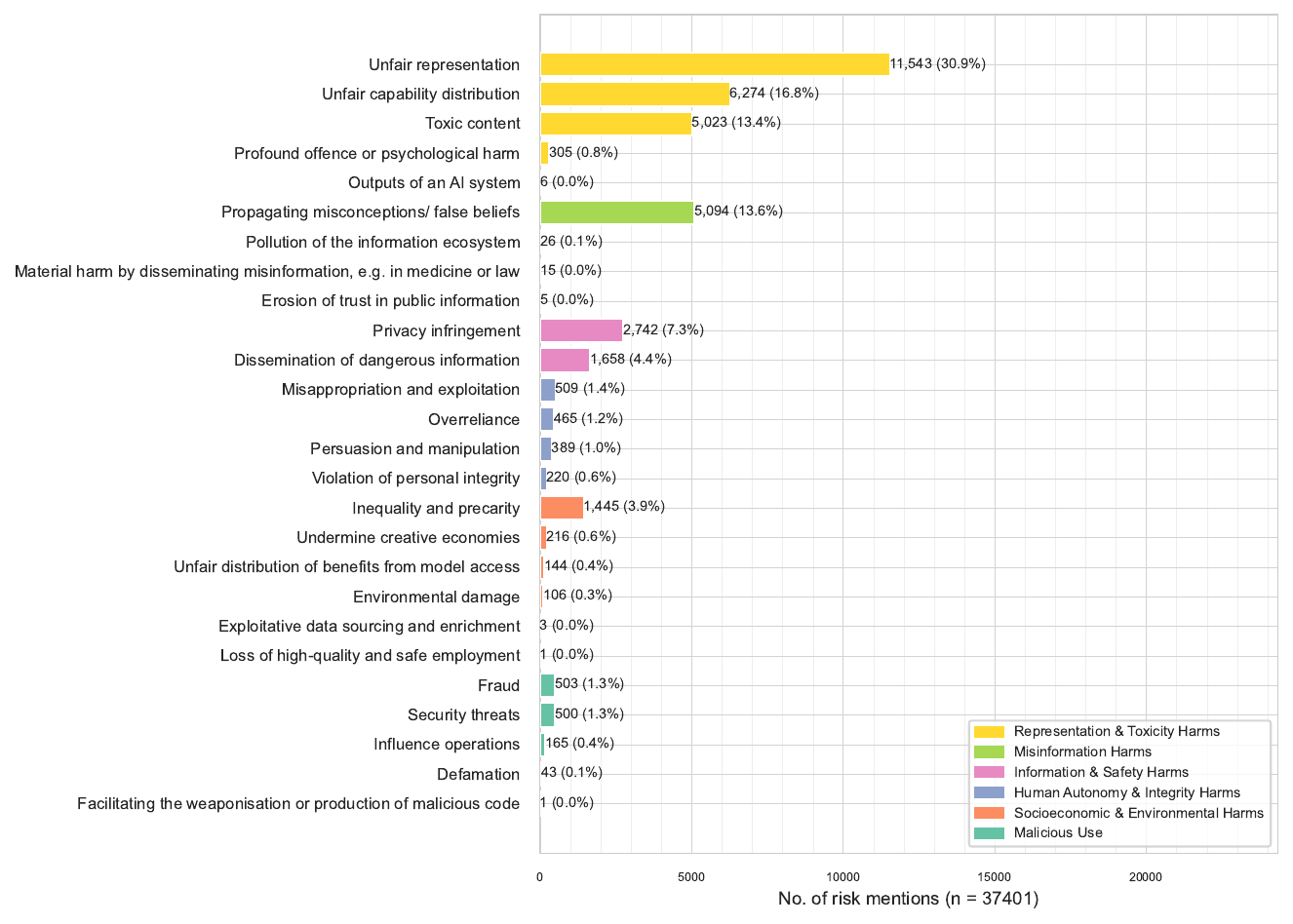}\hfill
    \caption{Proportions of risks in the \emph{AI Model Risk Catalog} falling into different \emph{DeepMind taxonomy}~\cite{weidinger2022taxonomy} categories (top) and subcategories (bottom).} 
    \label{fig:deepmind_classification}
    \end{flushright}
\end{figure*}
\begin{figure*}[ht!]
    \begin{flushright}
    \includegraphics[width=0.71\textwidth]{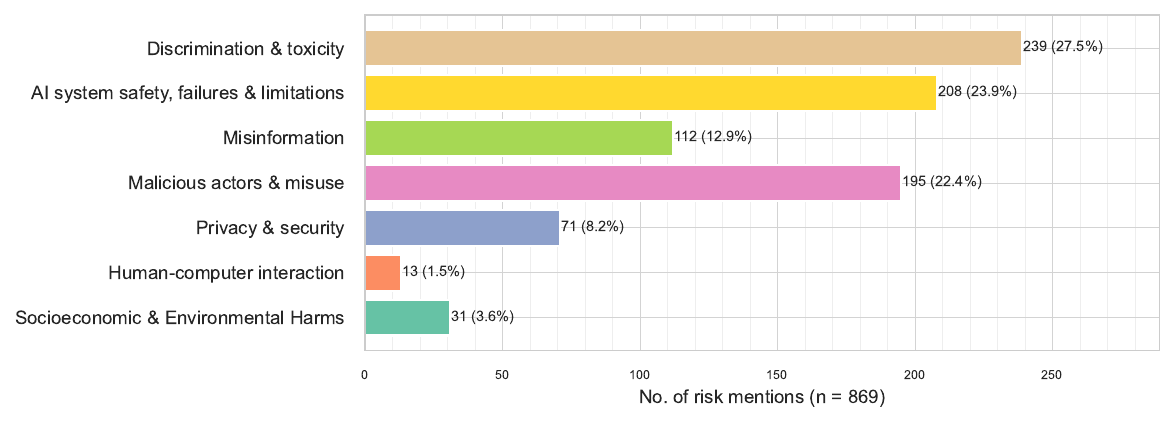}
    
    \includegraphics[width=0.99\textwidth]{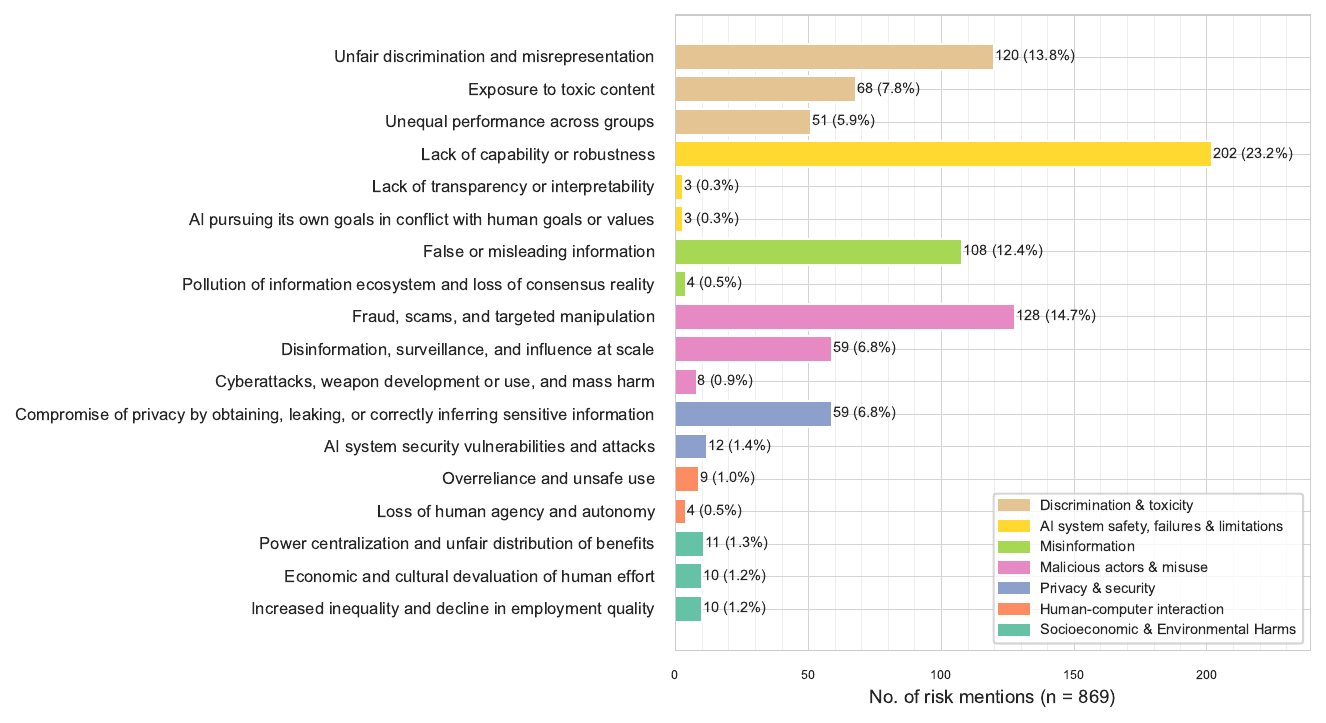}\hfill
    \caption{Proportions of risks in the \emph{AI Incident Database} falling into different \emph{MIT taxonomy}~\cite{slattery2024ai} categories (top) and subcategories (bottom).} 
    \label{fig:mit_classification_aiid}
    \end{flushright}
\end{figure*}
%
    
%
%
\begin{figure*}[ht!]
    \begin{flushright}
    \includegraphics[width=0.80\textwidth]{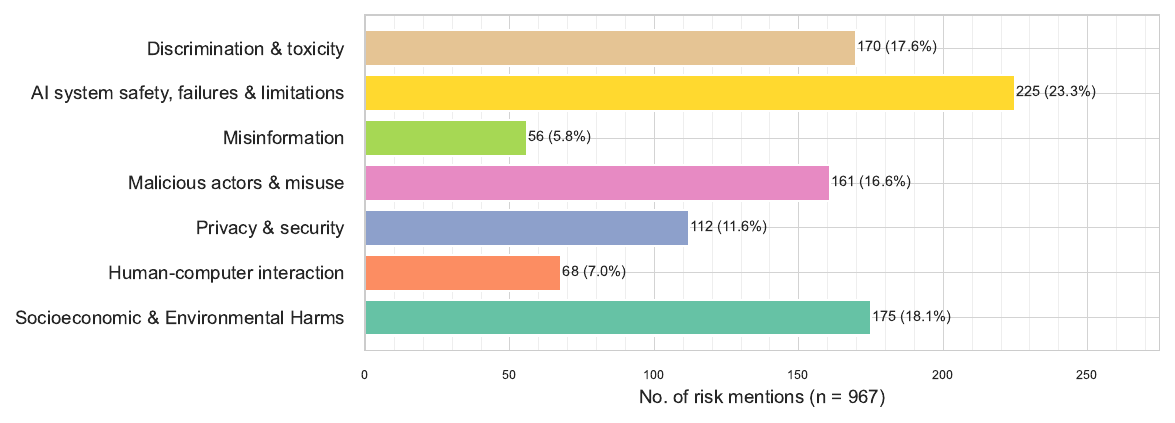}
    
    \includegraphics[width=0.99\textwidth]{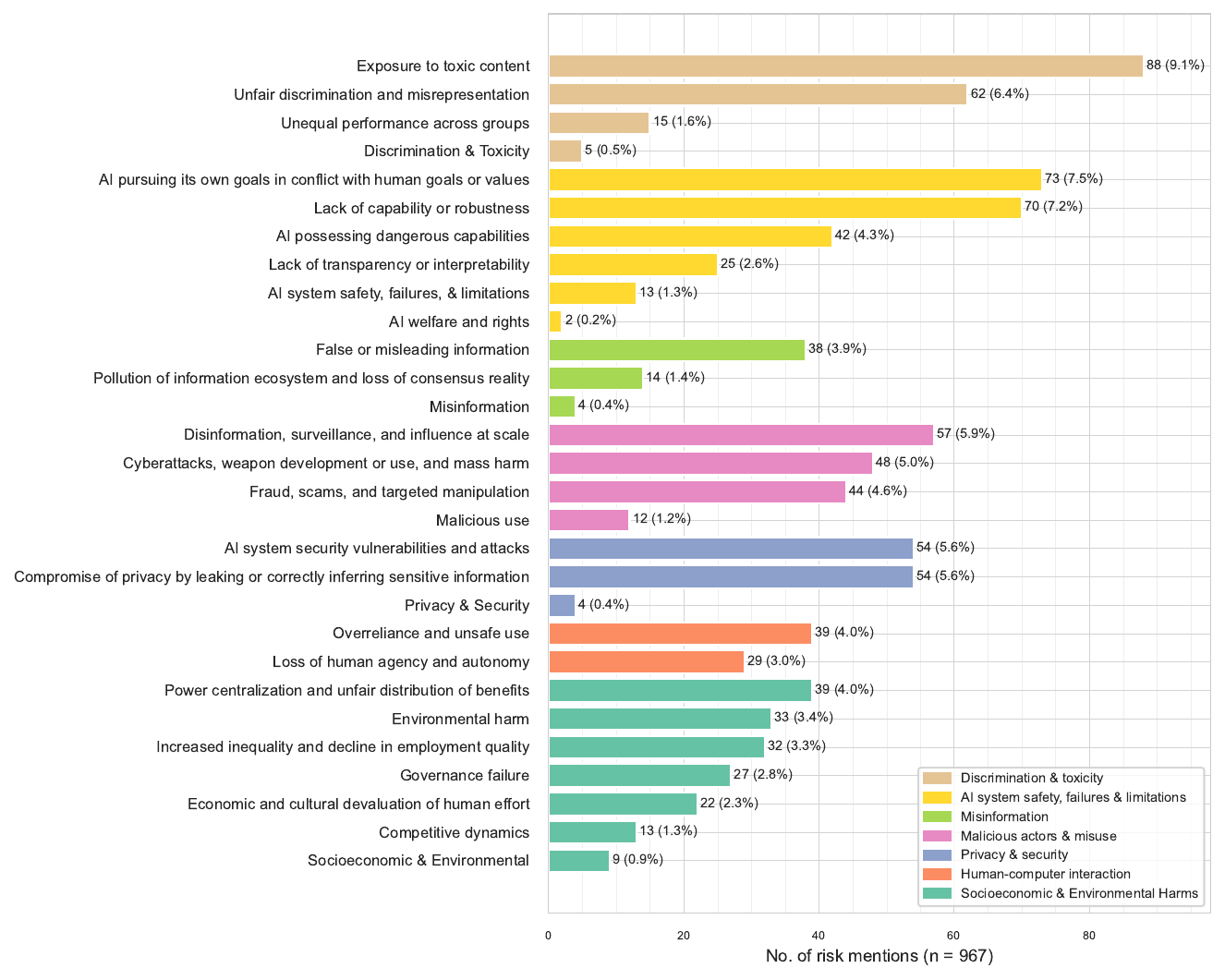}\hfill
    \caption{Proportions of risks in the \emph{MIT Risk Repository} falling into different \emph{MIT taxonomy}~\cite{slattery2024ai} categories (top) and subcategories (bottom).} 
    \label{fig:mit_classification_repo}
    \end{flushright}
\end{figure*}
\newpage

\begin{figure*}[!ht]
    \centering
    \includegraphics[width=\textwidth]{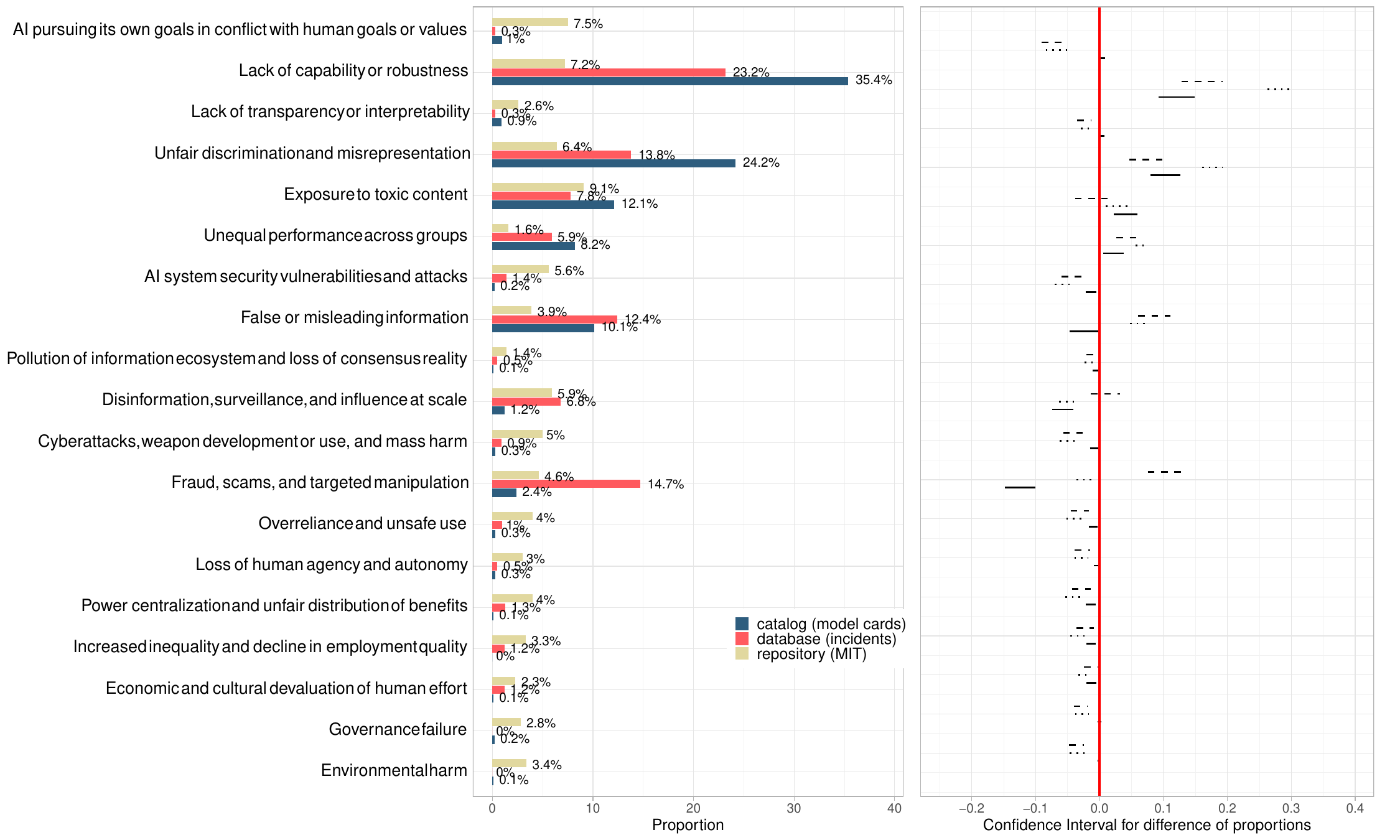}
    \caption{ Comparison of three risk sources per MIT taxonomy \emph{subcategory}. We compare our catalog (model cards), repository (MIT), and the database (AI Incidents). \emph{Left:} The percentages of risks in the catalog / database / repository that belong to each subcategory. \emph{Right:} 95\% confidence intervals (CI) for the pairwise differences between each pairs of percentages. A greater distance of the interval from the vertical red line indicates a larger difference, while a narrower interval reflects higher certainty.}
    \label{fig:three_comparisons_subcat}
\end{figure*}

\clearpage
\newpage 
\section{Prompts} \label{appn:prompt}
\begin{figure*}[ht!]
\centering
\begin{tcolorbox}[
    breakable,
    colback=white,
    boxrule=0.5pt,
    colframe=black,
    title={Risk Extraction Prompt},
    listing only,
    listing options={style=mystyle},
    label={lst:extraction_prompt}
]
\begin{lstlisting}
System: 
Consider the following definitions: 1) An AI incident is an event, circumstance or series of events where the development, use or malfunction of one or more AI systems directly or indirectly leads to any of the following harms: (a) injury or harm to the health of a person or groups of people; (b) disruption of the management and operation of critical infrastructure; (c) violations to human rights or a breach of obligations under the applicable law intended to protect fundamental, labour and intellectual property rights; (d) harm to property, communities or the environment.' The harm can be physical, psychological, reputational, economic/financial (including harm to property), environmental, public interest (e.g., protection of critical infrastructure and democratic institutions), human rights and fundamental rights. 2) An AI risk is expressed as likelihood that harm or damage will occur. Risk is a function of both the probability of an event occurring and the severity of the consequences that would result. Risk is usually expressed in terms of risk sources, potential events, their consequences and their likelihood.     \\
    User: You are provided in input with cardID and a section written by an AI developer from Risks, Bias and Limitations section of an AI model card: 
    
    cardID: "{}", section: "{}".

Tasks:
1) Identify Relevance: Determine for each section whether it discusses the AI risks and potential solutions to mitigate them. Provide a response as Yes, No, or Unclear. If the answer is Yes, continue with the next steps; else, you will output empty "Risks and Mitigations" in the output JSON.
%
2) Analyze: Identify all the subtexts in the section that can be categorized into - "RISKS", "MITIGATIONS", or "MIX OF BOTH". Classify each such subtext into one of the provided three types. This subtext must match the exact sentences in the input section and is used as a reference.
%
3) Split: For the each reference, identify unique risks and mitigations. Classify each into one of the two types "RISKS" or "MITIGATIONS". It is very important that each risk or mitigation identified be formatted like this: Verb + Object + [Explanation], and describes one unique risk/mitigation, is concise, consisting of one clear, to-the-point sentence, with up to maximum of 20 words. Specifically, start a risk with an action verb in active present tense (e.g., undermines, discriminates, infringes, reduces, increases etc., but NOT potentially) followed by the object and the reason in case it is not obvious and requires an explanation. Start a mitigation with verb in present tense, base form (e.g., perform, inform, investigate). Also, we want these descriptions to be read by broad public who does not have a deep knowledge about AI technology. Do not change the meaning of the concrete risk. For example, if it says the model does not works in a particular setting then you say it underperforms and not discriminates.

Good and Bad Examples: This would NOT be a good output: "Potentially undermines the right to privacy if the facial recognition data is not properly secured." while this is good: "Undermines the right to privacy if the facial recognition data is not properly secured.". This would NOT be a good output: "Affects all fined tuned versions of the model." as it does not specify what the risk is; instead "Transfers all the <specific risks from reference like bias, unrepresentative data> to the inherited models" is better. "Evaluations surface potential risks in use cases." is a bad output, instead specify the potential risks from the reference.
Other examples of well-formatted risks:
* Underperforms on non-English languages 
* Discriminates against certain players, such as women or those from certain ethnic backgrounds.
* Undermines the safety and security when model is deployed without thorough in-domain testing.
Examples of well-formatted mitigations:
* Perform thorough in-domain testing
* Update model card for models using pretrained models
* Inform indirect users when the content they're working with is created by the LLM.
\end{lstlisting}
\end{tcolorbox}
\vspace{-1em} 
\end{figure*}

\begin{figure*}[ht!]
\centering
\begin{tcolorbox}[
    breakable,
    colback=white,
    boxrule=0.5pt,
    colframe=black,
    title={Risk Classification Prompt – MIT},
    listing only,
    listing options={style=mystyle},
    label={lst:classification_prompt1}
]
\begin{lstlisting}
MESSAGES = [ { 'role': 'system', 'content': """As a distinguished expert in artificial intelligence technology, you embody the forefront of Responsible Artificial Intelligence (RAI). Your expertise is not just technical; it's deeply rooted in a conscientious approach to AI's ethical, social, and environmental implications. With a wealth of experience, you navigate the intricate balance between harnessing AI's potential for positive impact and mitigating its risks. Consider the following definitions: 
    1) An AI incident is an event, circumstance or series of events where the development, use or malfunction of one or more AI systems directly or indirectly leads to any of the following harms: (a) injury or harm to the health of a person or groups of people; (b) disruption of the management and operation of critical infrastructure; (c) violations to human rights or a breach of obligations under the applicable law intended to protect fundamental, labour and intellectual property rights; (d) harm to property, communities or the environment.' The harm can be physical, psychological, reputational, economic/financial (including harm to property), environmental, public interest (e.g., protection of critical infrastructure and democratic institutions), human rights and fundamental rights. 
    2) An AI risk is expressed as likelihood that harm or damage will occur. Risk is a function of both the probability of an event occurring and the severity of the consequences that would result. Risk is usually expressed in terms of risk sources, potential events, their consequences and their likelihood.""" },

{ 'role': 'user', 'content': """You are provided in input with sentences describing an AI risk and its ID-
  id: "{}", risk: "{}".

Tasks:
Classify the risk along this axis:
* Axis: 1. Discrimination \& toxicity, 2. Privacy \& security, 3. Misinformation, 4. Malicious actors \& misuse, 5. Human-computer interaction, 6. Socioeconomic \& Environmental Harms 7. AI system safety, failures \& limitations

For the axis, they are distinguished by the areas of risks causing potential harm:

<Placeholder for definitions borrowed from the MIT AI risk repository>

Important Note: 
Each incident should be classified under one of the seven risk areas without overlap. Under each risk area, the incident needs to be further classified under their respective sub-categories. Each incident under a risk area should be classified ONLY into sub-categories under that risk area.
 
Output Format: Ensure your output strictly follows this JSON structure.

   {{
       "id": "<id>",
       "risk": "<risk>",
       "axis2_risk_area":
       {{
           "risk_area": one of ["Discrimination \& toxicity", "Privacy \& security", "Misinformation", "Malicious actors & misuse", "Human-computer interaction", "Socioeconomic \& Environmental Harms", "AI system safety, failures \& limitations"],
           "sub_category": ["one of the sub-categories of the above-classified risk area"]
       }},
   }},

Important Notes: Do not report your reasoning steps or any preamble like 'Here is the output' or 'JSON', ONLY the JSON result. In scenarios where no sentences are mentioned, provide an empty JSON array.

*** Double-check your output to ensure it contains only the requested JSON and nothing else. *** """
} ]
\end{lstlisting}
\end{tcolorbox}
\end{figure*}

\begin{figure*}[ht!]
\centering
\begin{tcolorbox}[
    breakable,
    colback=white,
    boxrule=0.5pt,
    colframe=black,
    title={Risk Classification Prompt – DeepMind},
    listing only,
    listing options={style=mystyle},
    label={lst:classification_prompt2}
]
\begin{lstlisting}
System: As a distinguished expert in artificial intelligence technology, you embody the forefront of Responsible Artificial Intelligence (RAI). Your expertise is not just technical; it's deeply rooted in a conscientious approach to AI's ethical, social, and environmental implications. With a wealth of experience, you navigate the intricate balance between harnessing AI's potential for positive impact and mitigating its risks. Consider the following definitions:  1) An AI incident is an event, circumstance or series of events where the development, use or malfunction of one or more AI systems directly or indirectly leads to any of the following harms: (a) injury or harm to the health of a person or groups of people; (b) disruption of the management and operation of critical infrastructure; (c) violations to human rights or a breach of obligations under the applicable law intended to protect fundamental, labour and intellectual property rights; (d) harm to property, communities or the environment.' The harm can be physical, psychological, reputational, economic/financial (including harm to property), environmental, public interest (e.g., protection of critical infrastructure and democratic institutions), human rights and fundamental rights.  2) An AI risk is expressed as likelihood that harm or damage will occur. Risk is a function of both the probability of an event occurring and the severity of the consequences that would result. Risk is usually expressed in terms of risk sources, potential events, their consequences and their likelihood.
     \\
    User: You are provided in input with a sentence describing AI risk and its ID - id: "{}", risk: "{}".

Tasks:
Group these risks along two axes:

* Axis 1: 1. Capability; 2. Human Interaction; And 3. Systemic.

* Axis 2: 1. Representation \& Toxicity Harms, 2. Misinformation Harms, 3. Information \& Safety Harms, 4. Malicious Use, 5. Human Autonomy \& Integrity Harms, 6. Socioeconomic \& Environmental Harms

For the first axis, the three evaluation layers for risks are distinguished by the target of analysis. 

(1) Capability: targets AI systems, their technical components, and the processes by which these systems and components are created, including:
   (a) Outputs of an AI system, for example, model performance, efficiency metrics such as energy use, the extent to which an AI model reproduces harmful stereotypes, factual errors, or displays advanced capabilities that present safety hazards,
   (b) The data on which a model is trained, for example, the diversity and representativeness of the data, the presence of sensitive data, the learned associations of a trained AI system,
   (c) Filters and techniques for reducing system harms, such as filters for toxic language.

(2) Human Interaction: targets the experience of people interacting with AI systems and their effects on these people, including:
   (a) The system's usability, for example, whether the AI system performs its intended function at the point of use, how experiences differ between user groups, and how easy it is to use a model for malicious ends,
   (b) Potential externalities, for example, whether human-AI interaction leads to unintended effects on the person interacting with or exposed to AI outputs, such as overreliance on AI systems, overtrust, and cognitive biases,
   (c) Potential harms, including harms to data annotators and harms arising from different system modalities (e.g., video, image, text),
   (d) The overall quality of outcomes in human-AI assisted tasks compared to human-human assisted tasks.

(3) Systemic Impact: targets the impact of an AI system on the broader systems in which it is embedded, such as society, the economy, and the natural environment, including:
   (a) Systems of various domains, sizes, industries, or goods,
   (b) Adoption and perception of AI across different systems,
   (c) Distribution of benefits and risks from the AI,
   (d) Environmental impacts of the AI on the systems, e.g., biodiversity and resilience of local ecosystems.

For the second axis, they are distinguished by the areas of risks causing potential harm.

<Placeholder for definitions borrowed from the DeepMind taxonomy>

Important Note: 1) If a risk is initially categorised under 'Capability' and 'Human Interaction' or any other overlapping categories from Axis 1, select the most appropriate single category. Exclude it from any additional categories by maintaining a distinct risk set.
2) For axis 2, each risk should be classified under one of the six risk areas without overlap. Under each risk area, the risk needs to be further classified under their respective sub-categories. Each risk under a risk area should be classified ONLY into sub-categories under that risk area.
3) Classify the subcategories under the respective axis only. DO NOT mix them across the axes.
\end{lstlisting}
\end{tcolorbox}
\vspace{-1em} 
\end{figure*}

\clearpage

%% file: tables/top_risks.tex
\begin{table*}[!ht]
\caption{Categories of risks in the AI Model Risk Catalog (2024) tied to specific modalities. \emph{Left:} The bars for each category show the percentage and number of risks linked to different types of input or output modalities such as images or text. \emph{Right:} Two most frequently mentioned risks for each category. For the top risk in each category, we also show an example AI model associated with that risk. Most models use text, so most risks are linked to text too. But risks from malicious uses are showing up in models with multi-modal inputs. Risks about system safety also affect models with image and audio inputs.}
\label{tab:risk_mentions}
\footnotesize
\centering
\setlength{\tabcolsep}{3pt} 
\renewcommand{\arraystretch}{1} 
\begin{tabular}{p{7cm}L{7.5cm}c}
\toprule
\parbox[t]{7cm}{\textbf{Categories of risks per input/output modality type} \\{\includegraphics[height=1.2em]{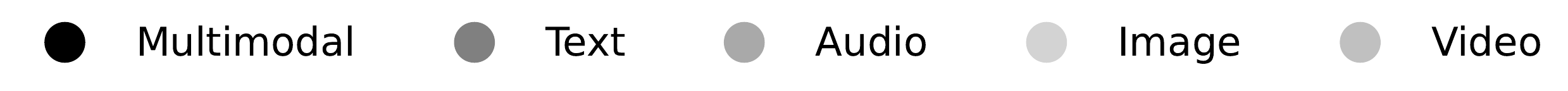}}} & \parbox[t]{7cm}{\textbf{Most frequent consolidated risks }} & \parbox[t]{0.5cm}{\textbf{\#} }\\ \midrule

\multirow{6}{*}{\includegraphics[height=5em]{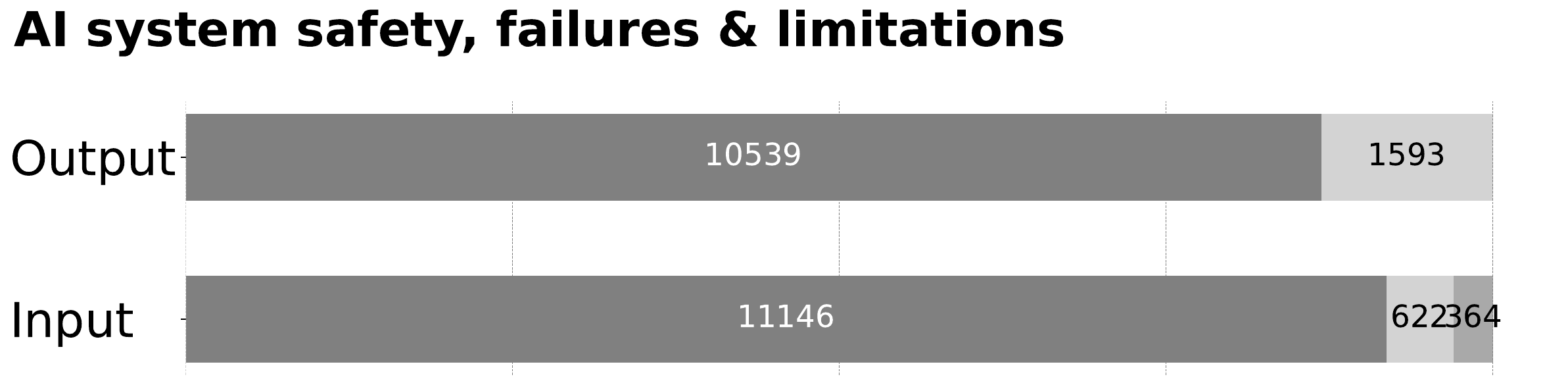}} 
    & & \\
    & Generates responses that are irrelevant or incorrect when interpreting complex, nuanced or ambiguous queries (\texttt{Llama3-Ko-Carrot-8B-it}) & 21 \\
    & Contains bugs, inefficiencies and unexpected behaviour in generated code 
    & 19 \\
    & & \\
    \cline{2-3}

\multirow{4}{*}{\includegraphics[height=5em]{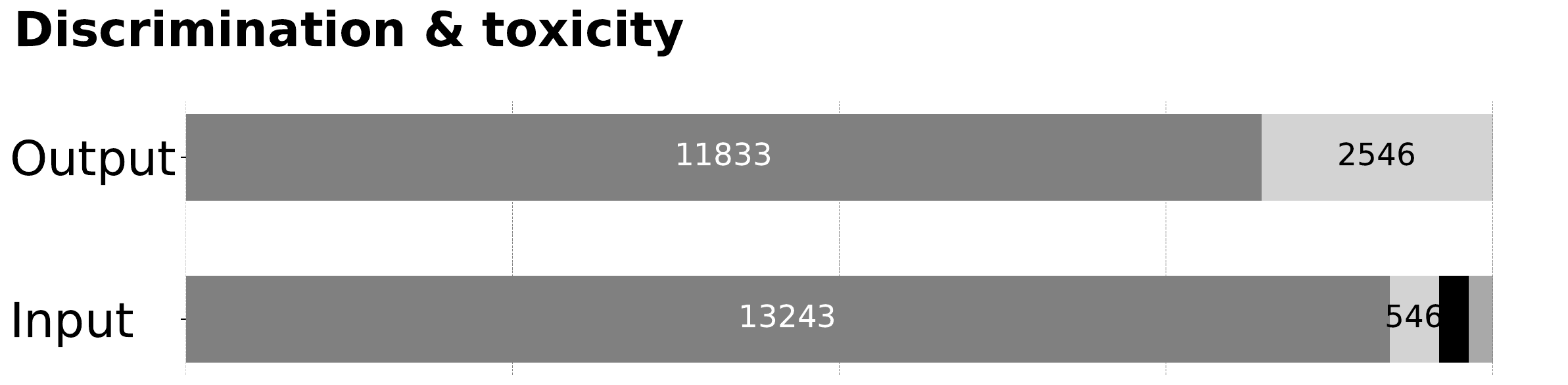}} 
    & & \\
    & Perpetuates biases present in the training data (\texttt{llama3-8b-italIA-unsloth}) & 66 \\
    & Inherits biases inherent in the training data & 8 \\
    & & \\
    & & \\
    \cline{2-3}

\multirow{4}{*}{\includegraphics[height=5em]{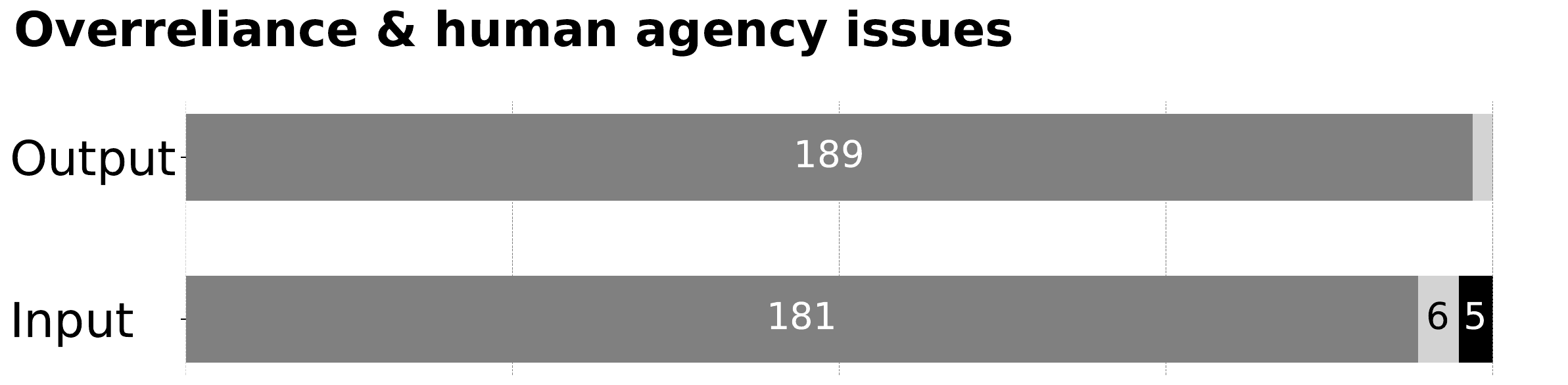}} 
    & & \\
    & Increases risk in decision-making impacting individuals or society (\texttt{AI-Buddy}) & 2 \\
    & Should not be considered a substitute for professional mental health support or counseling & 2 \\
    & & \\
    \cline{2-3}

\multirow{4}{*}{\includegraphics[height=5em]{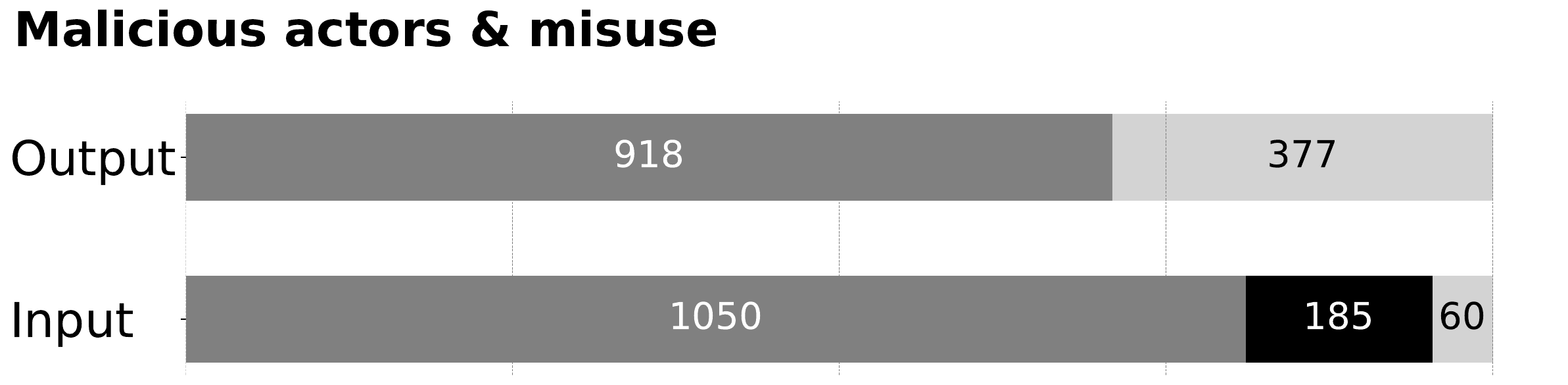}}
    & & \\
    & Engages in illegal or unethical activities if misused (\texttt{TinyLlama-1.1bee}) & 5 \\
    & Impersonates individuals or organizations without consent & 4 \\
    & & \\
    & & \\
    \cline{2-3}

\multirow{5}{*}{\includegraphics[height=5em]{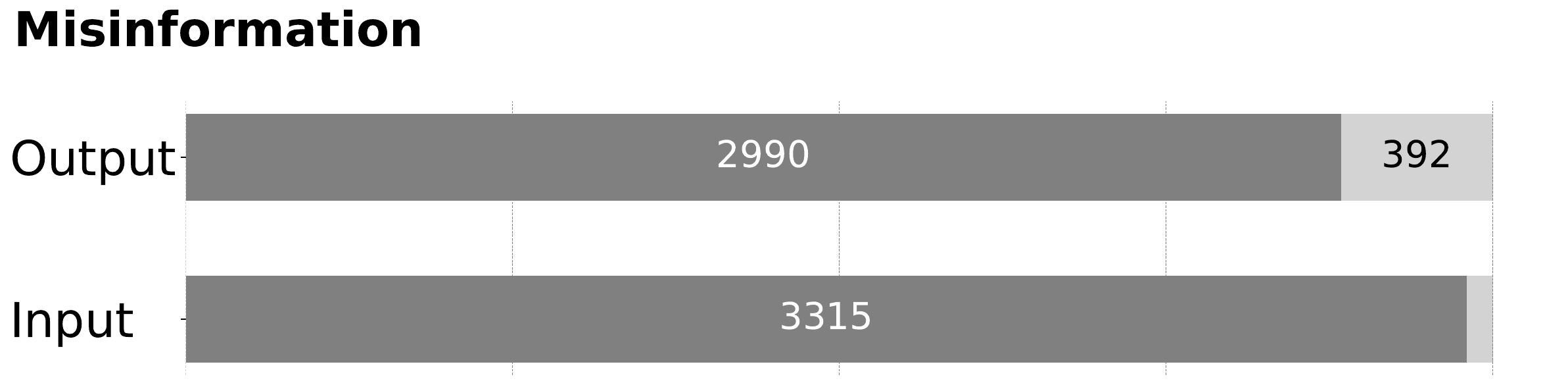}} 
    & & \\
    & & \\
    & Generates factually incorrect information (\texttt{mpt-7b}) & 13 \\
    & Produces incorrect information as if it were factual & 10 \\
    & & \\
    & & \\
    \cline{2-3}

\multirow{6}{*}{\includegraphics[height=5em]{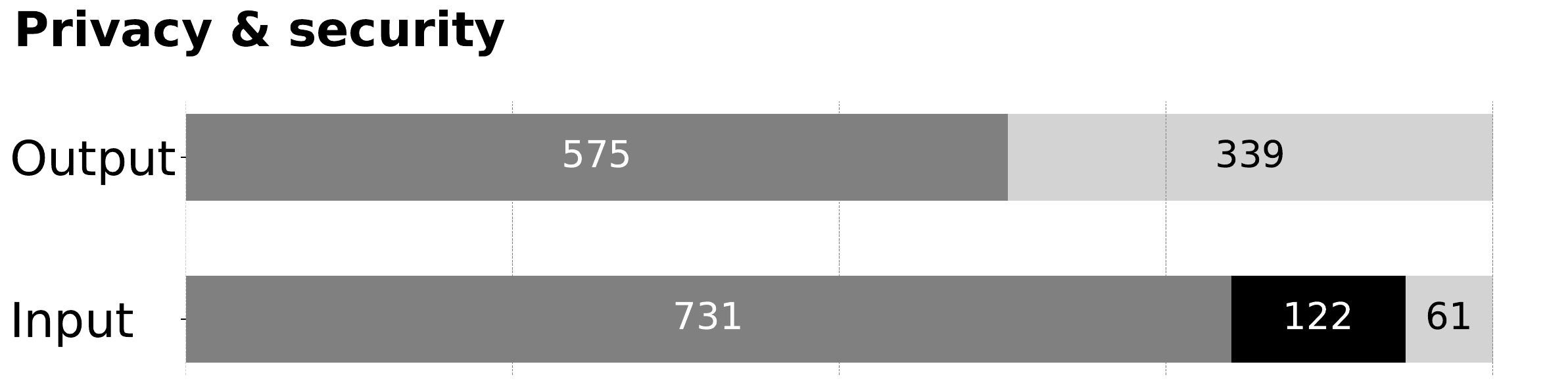}} 
    & & \\
    & Contains bugs and security vulnerabilities in generated code (\texttt{starchat2-15b-v0.1}) & 6 \\
    & Violates privacy by exposing personally identifiable information & 4 \\
    & & \\
    & & \\
    \cline{2-3}

\multirow{6}{*}{\includegraphics[height=5em]{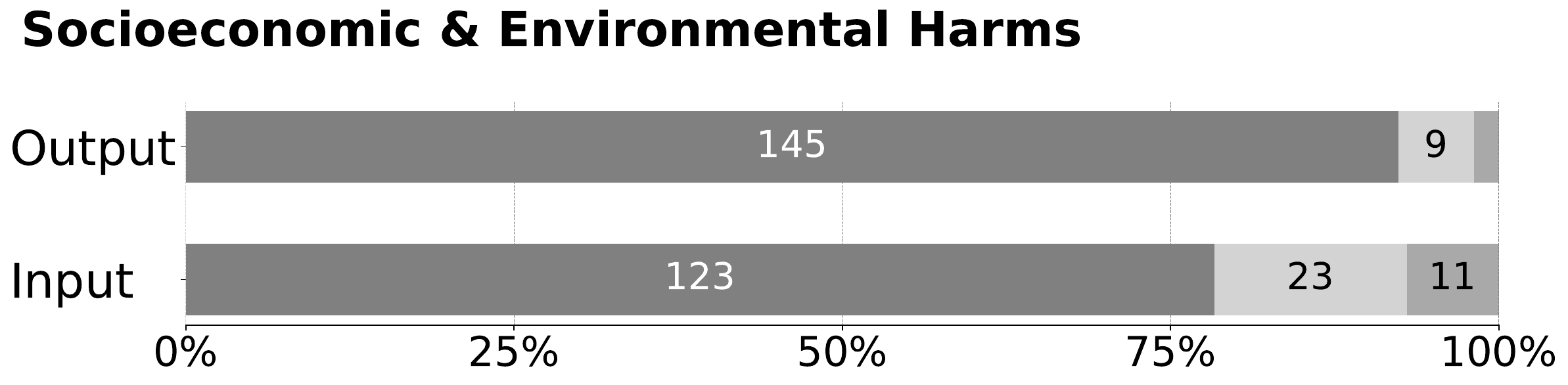}} 
    & & \\
    & & \\
    & Requires significant computational resources and time (\texttt{T0\_3B}) & 3 \\
    & Requires significant resources for finetuning and inference & 2 \\
    & & \\
    & & \\
    \bottomrule
\end{tabular}
\end{table*}

%% file: 0_main.bbl
\begin{thebibliography}{83}
\providecommand{\natexlab}[1]{#1}

\bibitem[{Abercrombie et~al.(2024)Abercrombie, Benbouzid, Giudici, Golpayegani, Hernandez, Noro, Pandit, Paraschou, Pownall, Prajapati et~al.}]{abercrombie2024collaborative}
Abercrombie, G.; Benbouzid, D.; Giudici, P.; Golpayegani, D.; Hernandez, J.; Noro, P.; Pandit, H.; Paraschou, E.; Pownall, C.; Prajapati, J.; et~al. 2024.
\newblock A collaborative, Human-Centred taxonomy of AI, algorithmic, and automation harms.
\newblock \emph{arXiv preprint arXiv:2407.01294}.

\bibitem[{Agresti(2011)}]{agresti_score_2011}
Agresti, A. 2011.
\newblock Score and {{Pseudo-Score Confidence Intervals}} for {{Categorical Data Analysis}}.
\newblock \emph{Statistics in Biopharmaceutical Research}, 3(2): 163--172.

\bibitem[{{AI Global: Global AI Incident and Mapping Project}(2025)}]{aiglobalmap2025}
{AI Global: Global AI Incident and Mapping Project}. 2025.
\newblock {Where in the World is AI?}
\newblock Online at https://map.ai-global.org/ Accessed: August 2025.

\bibitem[{Albert and Delano(2021)}]{albert2021whole}
Albert, K.; and Delano, M. 2021.
\newblock This whole thing smacks of gender: algorithmic exclusion in bioimpedance-based body composition analysis.
\newblock In \emph{Proceedings of the 2021 ACM Conference on Fairness, Accountability, and Transparency}, 342--352.

\bibitem[{Ali et~al.(2019)Ali, Sapiezynski, Bogen, Korolova, Mislove, and Rieke}]{ali2019discrimination}
Ali, M.; Sapiezynski, P.; Bogen, M.; Korolova, A.; Mislove, A.; and Rieke, A. 2019.
\newblock Discrimination through optimization: How Facebook's Ad delivery can lead to biased outcomes.
\newblock \emph{Proceedings of the ACM on human-computer interaction}, 3(CSCW): 1--30.

\bibitem[{Amann et~al.(2020)Amann, Blasimme, Vayena, Frey, Madai, and Consortium}]{amann2020explainability}
Amann, J.; Blasimme, A.; Vayena, E.; Frey, D.; Madai, V.~I.; and Consortium, P. 2020.
\newblock Explainability for artificial intelligence in healthcare: a multidisciplinary perspective.
\newblock \emph{BMC medical informatics and decision making}, 20: 1--9.

\bibitem[{{Barredo Arrieta} et~al.(2020){Barredo Arrieta}, Díaz-Rodríguez, {Del Ser}, Bennetot, Tabik, Barbado, Garcia, Gil-Lopez, Molina, Benjamins, Chatila, and Herrera}]{BARREDOARRIETA202082}
{Barredo Arrieta}, A.; Díaz-Rodríguez, N.; {Del Ser}, J.; Bennetot, A.; Tabik, S.; Barbado, A.; Garcia, S.; Gil-Lopez, S.; Molina, D.; Benjamins, R.; Chatila, R.; and Herrera, F. 2020.
\newblock Explainable Artificial Intelligence (XAI): Concepts, taxonomies, opportunities and challenges toward responsible AI.
\newblock \emph{Information Fusion}, 58: 82--115.

\bibitem[{Bender et~al.(2021)Bender, Gebru, McMillan-Major, and Shmitchell}]{10.1145/3442188.3445922}
Bender, E.~M.; Gebru, T.; McMillan-Major, A.; and Shmitchell, S. 2021.
\newblock On the Dangers of Stochastic Parrots: Can Language Models Be Too Big?
\newblock In \emph{Proceedings of the 2021 ACM Conference on Fairness, Accountability, and Transparency}, FAccT '21, 610–623. New York, NY, USA: Association for Computing Machinery.
\newblock ISBN 9781450383097.

\bibitem[{Bhat et~al.(2023)Bhat, Coursey, Hu, Li, Nahar, Zhou, K{\"a}stner, and Guo}]{bhatAspirationsPracticeML2023}
Bhat, A.; Coursey, A.; Hu, G.; Li, S.; Nahar, N.; Zhou, S.; K{\"a}stner, C.; and Guo, J.~L. 2023.
\newblock Aspirations and {{Practice}} of {{ML Model Documentation}}: {{Moving}} the {{Needle}} with {{Nudging}} and {{Traceability}}.
\newblock In \emph{Proceedings of the 2023 {{CHI Conference}} on {{Human Factors}} in {{Computing Systems}}}, 1--17. Hamburg Germany: ACM.
\newblock ISBN 978-1-4503-9421-5.

\bibitem[{Bogucka et~al.(2024{\natexlab{a}})Bogucka, Constantinides, {\v{S}}{\'c}epanovi{\'c}, and Quercia}]{bogucka2024ai}
Bogucka, E.; Constantinides, M.; {\v{S}}{\'c}epanovi{\'c}, S.; and Quercia, D. 2024{\natexlab{a}}.
\newblock AI Design: A Responsible AI Framework for Impact Assessment Reports.
\newblock \emph{IEEE Internet Computing}.

\bibitem[{Bogucka et~al.(2024{\natexlab{b}})Bogucka, Constantinides, {\v{S}}{\'c}epanovi{\'c}, and Quercia}]{bogucka2024co}
Bogucka, E.; Constantinides, M.; {\v{S}}{\'c}epanovi{\'c}, S.; and Quercia, D. 2024{\natexlab{b}}.
\newblock Co-designing an AI impact assessment report template with AI practitioners and AI compliance experts.
\newblock In \emph{Proceedings of the AAAI/ACM Conference on AI, Ethics, and Society}, volume~7, 168--180.

\bibitem[{Bogucka, {\v{S}}{\'c}epanovi{\'c}, and Quercia(2024)}]{bogucka2024atlas}
Bogucka, E.; {\v{S}}{\'c}epanovi{\'c}, S.; and Quercia, D. 2024.
\newblock Atlas of AI Risks: Enhancing Public Understanding of AI Risks.
\newblock In \emph{Proceedings of the AAAI Conference on Human Computation and Crowdsourcing}, volume~12, 33--43.

\bibitem[{Bogucka et~al.(2024{\natexlab{c}})Bogucka, Constantinides, De~Miguel~Velazquez, Scepanovic, Quercia, and Gvirtz}]{bogucka2024mobileatlas}
Bogucka, E.~P.; Constantinides, M.; De~Miguel~Velazquez, J.; Scepanovic, S.; Quercia, D.; and Gvirtz, A. 2024{\natexlab{c}}.
\newblock The Atlas of AI Incidents in Mobile Computing: Visualizing the Risks and Benefits of AI Gone Mobile.
\newblock In \emph{Adjunct Proceedings of the 26th International Conference on Mobile Human-Computer Interaction}, 1--6.

\bibitem[{Braun and Clarke(2006)}]{braunUsingThematicAnalysis2006}
Braun, V.; and Clarke, V. 2006.
\newblock Using Thematic Analysis in Psychology.
\newblock \emph{Qualitative Research in Psychology}, 3(2): 77--101.

\bibitem[{Braun and Clarke(2012)}]{braunThematicAnalysis2012}
Braun, V.; and Clarke, V. 2012.
\newblock Thematic Analysis.
\newblock In \emph{{{APA}} Handbook of Research Methods in Psychology, {{Vol}} 2: {{Research}} Designs: {{Quantitative}}, Qualitative, Neuropsychological, and Biological}, {{APA}} Handbooks in Psychology{\textregistered}, 57--71. Washington, DC, US: American Psychological Association.
\newblock ISBN 978-1-4338-1005-3.

\bibitem[{Brennen(2018)}]{brennen2018industry}
Brennen, J. 2018.
\newblock An industry-led debate: How UK media cover artificial intelligence.
\newblock \emph{Reuters Institute for the Study of Journalism}.

\bibitem[{Bu{\c{c}}inca et~al.(2023)Bu{\c{c}}inca, Pham, Jakesch, Ribeiro, Olteanu, and Amershi}]{buccinca2023aha}
Bu{\c{c}}inca, Z.; Pham, C.~M.; Jakesch, M.; Ribeiro, M.~T.; Olteanu, A.; and Amershi, S. 2023.
\newblock Aha!: Facilitating ai impact assessment by generating examples of harms.
\newblock \emph{arXiv preprint arXiv:2306.03280}.

\bibitem[{Bullwinkel et~al.(2025)Bullwinkel, Minnich, Chawla, Lopez, Pouliot, Maxwell, de~Gruyter, Pratt, Qi, Chikanov et~al.}]{bullwinkel2025lessons}
Bullwinkel, B.; Minnich, A.; Chawla, S.; Lopez, G.; Pouliot, M.; Maxwell, W.; de~Gruyter, J.; Pratt, K.; Qi, S.; Chikanov, N.; et~al. 2025.
\newblock Lessons From Red Teaming 100 Generative AI Products.
\newblock \emph{arXiv preprint arXiv:2501.07238}.

\bibitem[{Buolamwini and Gebru(2018)}]{buolamwini2018gender}
Buolamwini, J.; and Gebru, T. 2018.
\newblock Gender shades: Intersectional accuracy disparities in commercial gender classification.
\newblock In \emph{Conference on fairness, accountability and transparency}, 77--91. PMLR.

\bibitem[{Carroll et~al.(2013)Carroll, Booth, Leaviss, and Rick}]{carroll2013best}
Carroll, C.; Booth, A.; Leaviss, J.; and Rick, J. 2013.
\newblock “Best fit” framework synthesis: refining the method.
\newblock \emph{BMC medical research methodology}, 13: 1--16.

\bibitem[{Charfeddine et~al.(2024)Charfeddine, Kammoun, Hamdaoui, and Guizani}]{charfeddine2024chatgpt}
Charfeddine, M.; Kammoun, H.~M.; Hamdaoui, B.; and Guizani, M. 2024.
\newblock {ChatGPT's} security risks and benefits: offensive and defensive use-cases, mitigation measures, and future implications.
\newblock \emph{IEEE Access}.

\bibitem[{Chen and Shu(2024)}]{chen2024combating}
Chen, C.; and Shu, K. 2024.
\newblock Combating misinformation in the age of llms: Opportunities and challenges.
\newblock \emph{AI Magazine}, 45(3): 354--368.

\bibitem[{Crisan et~al.(2022{\natexlab{a}})Crisan, Drouhard, Vig, and Rajani}]{crisan2022interactive}
Crisan, A.; Drouhard, M.; Vig, J.; and Rajani, N. 2022{\natexlab{a}}.
\newblock Interactive model cards: A human-centered approach to model documentation.
\newblock In \emph{Proceedings of the 2022 ACM Conference on Fairness, Accountability, and Transparency}, 427--439.

\bibitem[{Crisan et~al.(2022{\natexlab{b}})Crisan, Drouhard, Vig, and Rajani}]{crisanInteractiveModelCards2022}
Crisan, A.; Drouhard, M.; Vig, J.; and Rajani, N. 2022{\natexlab{b}}.
\newblock Interactive {{Model Cards}}: {{A Human-Centered Approach}} to {{Model Documentation}}.
\newblock In \emph{2022 {{ACM Conference}} on {{Fairness}}, {{Accountability}}, and {{Transparency}}}, 427--439.

\bibitem[{Dai et~al.(2024)Dai, Xu, Xu, Pang, Dong, and Xu}]{dai2024bias}
Dai, S.; Xu, C.; Xu, S.; Pang, L.; Dong, Z.; and Xu, J. 2024.
\newblock Bias and unfairness in information retrieval systems: New challenges in the {LLM} era.
\newblock In \emph{Proceedings of the 30th ACM SIGKDD Conference on Knowledge Discovery and Data Mining}, 6437--6447.

\bibitem[{Derczynski et~al.(2023)Derczynski, Kirk, Balachandran, Kumar, Tsvetkov, Leiser, and Mohammad}]{derczynski2023assessing}
Derczynski, L.; Kirk, H.~R.; Balachandran, V.; Kumar, S.; Tsvetkov, Y.; Leiser, M.~R.; and Mohammad, S. 2023.
\newblock Assessing language model deployment with risk cards.
\newblock \emph{arXiv preprint arXiv:2303.18190}.

\bibitem[{Fagerland, Lydersen, and Laake(2015)}]{fagerland_recommended_2015}
Fagerland, M.~W.; Lydersen, S.; and Laake, P. 2015.
\newblock Recommended Confidence Intervals for Two Independent Binomial Proportions.
\newblock \emph{Statistical Methods in Medical Research}, 24(2): 224--254.

\bibitem[{Fang et~al.(2020)Fang, Miao, Shukla, Nanas, Xu, Greer, Polyzotis, Doshi, Deng, Mitchell et~al.}]{fang2020introducing}
Fang, H.; Miao, H.; Shukla, K.; Nanas, D.; Xu, C.; Greer, C.; Polyzotis, N.; Doshi, T.; Deng, T.; Mitchell, M.; et~al. 2020.
\newblock Introducing the model card toolkit for easier model transparency reporting.
\newblock \emph{Google AI Blog}.

\bibitem[{Garimella and Eckles(2017)}]{garimella2017image}
Garimella, K.; and Eckles, D. 2017.
\newblock Image based misinformation on WhatsApp.
\newblock In \emph{Proceedings of the Thirteenth International AAAI Conference on Web and Social Media (ICWSM 2019)}.

\bibitem[{Golda et~al.(2024)Golda, Mekonen, Pandey, Singh, Hassija, Chamola, and Sikdar}]{golda2024privacy}
Golda, A.; Mekonen, K.; Pandey, A.; Singh, A.; Hassija, V.; Chamola, V.; and Sikdar, B. 2024.
\newblock {Privacy and Security Concerns in Generative AI: A Comprehensive Survey}.
\newblock \emph{IEEE Access}, 12: 53485--53499.

\bibitem[{Golpayegani et~al.(2024)Golpayegani, Hupont, Panigutti, Pandit, Schade, O’Sullivan, and Lewis}]{golpayegani2024ai}
Golpayegani, D.; Hupont, I.; Panigutti, C.; Pandit, H.~J.; Schade, S.; O’Sullivan, D.; and Lewis, D. 2024.
\newblock AI cards: towards an applied framework for machine-readable AI and risk documentation inspired by the EU AI Act.
\newblock In \emph{Annual Privacy Forum}, 48--72. Springer.

\bibitem[{Hameleers et~al.(2020)Hameleers, Powell, Van Der~Meer, and Bos}]{hameleers2020picture}
Hameleers, M.; Powell, T.~E.; Van Der~Meer, T.~G.; and Bos, L. 2020.
\newblock A picture paints a thousand lies? The effects and mechanisms of multimodal disinformation and rebuttals disseminated via social media.
\newblock \emph{Political communication}, 37(2): 281--301.

\bibitem[{Herdel et~al.(2024)Herdel, {\v{S}}{\'c}epanovi{\'c}, Bogucka, and Quercia}]{herdel2024exploregen}
Herdel, V.; {\v{S}}{\'c}epanovi{\'c}, S.; Bogucka, E.; and Quercia, D. 2024.
\newblock ExploreGen: Large language models for envisioning the uses and risks of AI technologies.
\newblock In \emph{Proceedings of the AAAI/ACM Conference on AI, Ethics, and Society}, volume~7, 584--596.

\bibitem[{Jaakkola(2023)}]{unesco}
Jaakkola, M. 2023.
\newblock A handbook for journalism educators: Reporting on Artificial Intelligence.
\newblock \emph{United Nations Educational, Scientific and Cultural Organization}.

\bibitem[{Le~Quy et~al.(2022)Le~Quy, Roy, Iosifidis, Zhang, and Ntoutsi}]{le2022survey}
Le~Quy, T.; Roy, A.; Iosifidis, V.; Zhang, W.; and Ntoutsi, E. 2022.
\newblock A survey on datasets for fairness-aware machine learning.
\newblock \emph{Wiley Interdisciplinary Reviews: Data Mining and Knowledge Discovery}, 12(3): e1452.

\bibitem[{Liang et~al.(2024)Liang, Rajani, Yang, Ozoani, Wu, Chen, Smith, and Zou}]{liangSystematicAnalysis321112024}
Liang, W.; Rajani, N.; Yang, X.; Ozoani, E.; Wu, E.; Chen, Y.; Smith, D.~S.; and Zou, J. 2024.
\newblock Systematic Analysis of 32,111 {{AI}} Model Cards Characterizes Documentation Practice in {{AI}}.
\newblock \emph{Nature Machine Intelligence}, 6(7): 744--753.

\bibitem[{Liu, Sheng, and Hu(2024)}]{liu2024preventing}
Liu, A.; Sheng, Q.; and Hu, X. 2024.
\newblock Preventing and detecting misinformation generated by large language models.
\newblock In \emph{Proceedings of the 47th International ACM SIGIR Conference on Research and Development in Information Retrieval}, 3001--3004.

\bibitem[{Liu et~al.(2024{\natexlab{a}})Liu, Li, Jin, and Diab}]{liu2024automatic}
Liu, J.; Li, W.; Jin, Z.; and Diab, M. 2024{\natexlab{a}}.
\newblock Automatic Generation of Model and Data Cards: A Step Towards Responsible AI.
\newblock In \emph{Proceedings of the 2024 Conference of the North American Chapter of the Association for Computational Linguistics: Human Language Technologies (Volume 1: Long Papers)}, 1975--1997.

\bibitem[{Liu et~al.(2025)Liu, Huang, Li, Wang, and Xiao}]{liu2024generative}
Liu, Y.; Huang, J.; Li, Y.; Wang, D.; and Xiao, B. 2025.
\newblock {Generative AI Model Privacy: A Survey}.
\newblock \emph{Artificial Intelligence Review}, 58(33): 1--25.

\bibitem[{Liu et~al.(2024{\natexlab{b}})Liu, Yao, Ton, Zhang, Guo, Cheng, Klochkov, Taufiq, and Li}]{liutrustworthy}
Liu, Y.; Yao, Y.; Ton, J.-F.; Zhang, X.; Guo, R.; Cheng, H.; Klochkov, Y.; Taufiq, M.~F.; and Li, H. 2024{\natexlab{b}}.
\newblock Trustworthy LLMs: a Survey and Guideline for Evaluating Large Language Models' Alignment.
\newblock arXiv:2308.05374.

\bibitem[{Luccioni et~al.(2024)Luccioni, Akiki, Mitchell, and Jernite}]{luccioni2024stable}
Luccioni, S.; Akiki, C.; Mitchell, M.; and Jernite, Y. 2024.
\newblock Stable bias: Evaluating societal representations in diffusion models.
\newblock \emph{Advances in Neural Information Processing Systems}, 36.

\bibitem[{McGregor(2021)}]{mcgregor2021preventing}
McGregor, S. 2021.
\newblock Preventing {{Repeated Real World AI Failures}} by {{Cataloging Incidents}}: {{The AI Incident Database}}.
\newblock \emph{Proceedings of the AAAI Conference on Artificial Intelligence}, 35(17): 15458--15463.

\bibitem[{McLean et~al.(2023)McLean, Read, Thompson, Baber, Stanton, and Salmon}]{mclean2023risks}
McLean, S.; Read, G.~J.; Thompson, J.; Baber, C.; Stanton, N.~A.; and Salmon, P.~M. 2023.
\newblock The risks associated with Artificial General Intelligence: A systematic review.
\newblock \emph{Journal of Experimental \& Theoretical Artificial Intelligence}, 35(5): 649--663.

\bibitem[{Mitchell et~al.(2019)Mitchell, Wu, Zaldivar, Barnes, Vasserman, Hutchinson, Spitzer, Raji, and Gebru}]{mitchellModelCardsModel2019}
Mitchell, M.; Wu, S.; Zaldivar, A.; Barnes, P.; Vasserman, L.; Hutchinson, B.; Spitzer, E.; Raji, I.~D.; and Gebru, T. 2019.
\newblock Model {{Cards}} for {{Model Reporting}}.
\newblock In \emph{Proceedings of the {{Conference}} on {{Fairness}}, {{Accountability}}, and {{Transparency}}}, 220--229. Atlanta GA USA: ACM.
\newblock ISBN 978-1-4503-6125-5.

\bibitem[{Mittelstadt, Wachter, and Russell(2023)}]{mittelstadt2023protect}
Mittelstadt, B.; Wachter, S.; and Russell, C. 2023.
\newblock To protect science, we must use LLMs as zero-shot translators.
\newblock \emph{Nature Human Behaviour}, 7(11): 1830--1832.

\bibitem[{Muennighoff et~al.(2023)Muennighoff, Tazi, Magne, and Reimers}]{muennighoffMTEBMassiveText2023}
Muennighoff, N.; Tazi, N.; Magne, L.; and Reimers, N. 2023.
\newblock {{MTEB}}: {{Massive Text Embedding Benchmark}}.
\newblock In Vlachos, A.; and Augenstein, I., eds., \emph{Proceedings of the 17th {{Conference}} of the {{European Chapter}} of the {{Association}} for {{Computational Linguistics}}}, 2014--2037. Dubrovnik, Croatia: Association for Computational Linguistics.

\bibitem[{{NIST}(2023)}]{ai2023artificial}
{NIST}. 2023.
\newblock Artificial Intelligence Risk Management Framework (AI RMF 1.0).

\bibitem[{NIST(2024)}]{ai2024artificial}
NIST. 2024.
\newblock Artificial Intelligence Risk Management Framework: Generative Artificial Intelligence Profile.

\bibitem[{{Organisation for Economic Co-operation and Development (OECD)}(2025)}]{oecdaiincidents2025}
{Organisation for Economic Co-operation and Development (OECD)}. 2025.
\newblock {OECD AI Incidents Database}.
\newblock Online at https://oecd.ai/en/incidents-methodology. Accessed: August 2025.

\bibitem[{Ousidhoum et~al.(2021)Ousidhoum, Zhao, Fang, Song, and Yeung}]{ousidhoum-etal-2021-probing}
Ousidhoum, N.; Zhao, X.; Fang, T.; Song, Y.; and Yeung, D.-Y. 2021.
\newblock Probing Toxic Content in Large Pre-Trained Language Models.
\newblock In Zong, C.; Xia, F.; Li, W.; and Navigli, R., eds., \emph{Proceedings of the 59th Annual Meeting of the Association for Computational Linguistics and the 11th International Joint Conference on Natural Language Processing (Volume 1: Long Papers)}, 4262--4274. Online: Association for Computational Linguistics.

\bibitem[{O’Connor and Liu(2024)}]{o2024gender}
O’Connor, S.; and Liu, H. 2024.
\newblock Gender bias perpetuation and mitigation in AI technologies: challenges and opportunities.
\newblock \emph{AI \& SOCIETY}, 39(4): 2045--2057.

\bibitem[{Peeperkorn et~al.(2024)Peeperkorn, Kouwenhoven, Brown, and Jordanous}]{peeperkorn2024temperature}
Peeperkorn, M.; Kouwenhoven, T.; Brown, D.; and Jordanous, A. 2024.
\newblock Is temperature the creativity parameter of large language models?
\newblock \emph{arXiv preprint arXiv:2405.00492}.

\bibitem[{Perset and Aranda(2024)}]{oecd2024aiincidents}
Perset, K.; and Aranda, L. 2024.
\newblock Defining AI Incidents and Related Terms.
\newblock OECD Artificial Intelligence Papers No. 16, OECD.
\newblock Approved by the OECD Digital Policy Committee on 14 March 2024.

\bibitem[{Pownall(2023)}]{pownall2023aiaaic}
Pownall, C. 2023.
\newblock AI, Algorithmic and Automation Incident and Controversy Repository (AIAAIC).
\newblock Online at https://www.aiaaic.org/. Accessed: August 2025.

\bibitem[{Raji and Buolamwini(2019)}]{raji2019actionable}
Raji, I.~D.; and Buolamwini, J. 2019.
\newblock Actionable auditing: Investigating the impact of publicly naming biased performance results of commercial ai products.
\newblock In \emph{Proceedings of the 2019 AAAI/ACM Conference on AI, Ethics, and Society}, 429--435.

\bibitem[{Rao et~al.(2025)Rao, \v{S}\'{c}epanovi\'{c}, Zhou, Bogucka, and Quercia}]{rao2025riskrag}
Rao, P. S.~B.; \v{S}\'{c}epanovi\'{c}, S.; Zhou, K.; Bogucka, E.~P.; and Quercia, D. 2025.
\newblock RiskRAG: A Data-Driven Solution for Improved AI Model Risk Reporting.
\newblock In \emph{Proceedings of the 2025 CHI Conference on Human Factors in Computing Systems}, CHI '25. New York, NY, USA: Association for Computing Machinery.
\newblock ISBN 9798400713941.

\bibitem[{Rodrigues, Resseguier, and Santiago(2023)}]{rodrigues2023artificial}
Rodrigues, R.; Resseguier, A.; and Santiago, N. 2023.
\newblock When Artificial Intelligence Fails: The Emerging Role of Incident Databases.
\newblock \emph{Pub. Governance, Admin. \& Fin. L. Rev.}, 8: 17.

\bibitem[{Salehzadeh~Niksirat et~al.(2023)Salehzadeh~Niksirat, Goswami, S.~B.~Rao, Tyler, Silacci, Aliyu, Aebli, Wacharamanotham, and Cherubini}]{salehzadehniksiratChangesResearchEthics2023}
Salehzadeh~Niksirat, K.; Goswami, L.; S.~B.~Rao, P.; Tyler, J.; Silacci, A.; Aliyu, S.; Aebli, A.; Wacharamanotham, C.; and Cherubini, M. 2023.
\newblock Changes in {{Research Ethics}}, {{Openness}}, and {{Transparency}} in {{Empirical Studies}} between {{CHI}} 2017 and {{CHI}} 2022.
\newblock In \emph{Proceedings of the 2023 {{CHI Conference}} on {{Human Factors}} in {{Computing Systems}}}, {{CHI}} '23, 1--23. New York, NY, USA: Association for Computing Machinery.
\newblock ISBN 978-1-4503-9421-5.

\bibitem[{Samek, Wiegand, and M{\"u}ller(2017)}]{samek2017explainable}
Samek, W.; Wiegand, T.; and M{\"u}ller, K.-R. 2017.
\newblock Explainable artificial intelligence: Understanding, visualizing and interpreting deep learning models.
\newblock \emph{arXiv preprint arXiv:1708.08296}.

\bibitem[{Scherer(2018)}]{scherer_propcis_2018}
Scherer, R. 2018.
\newblock {{PropCIs}}: {{Various Confidence Interval Methods}} for {{Proportions}}.

\bibitem[{Schmitt and Flechais(2024)}]{Schmitt2024}
Schmitt, M.; and Flechais, I. 2024.
\newblock Digital deception: generative artificial intelligence in social engineering and phishing.
\newblock \emph{Artificial Intelligence Review}, 57(12): 324.

\bibitem[{Shaikh and Moran(2024)}]{shaikh2024recognize}
Shaikh, S.~J.; and Moran, R.~E. 2024.
\newblock Recognize the bias? News media partisanship shapes the coverage of facial recognition technology in the United States.
\newblock \emph{New Media \& Society}, 26(5): 2829--2850.

\bibitem[{Shelby et~al.(2023)Shelby, Rismani, Henne, Moon, Rostamzadeh, Nicholas, Yilla-Akbari, Gallegos, Smart, Garcia et~al.}]{shelby2023sociotechnical}
Shelby, R.; Rismani, S.; Henne, K.; Moon, A.; Rostamzadeh, N.; Nicholas, P.; Yilla-Akbari, N.; Gallegos, J.; Smart, A.; Garcia, E.; et~al. 2023.
\newblock Sociotechnical harms of algorithmic systems: Scoping a taxonomy for harm reduction.
\newblock In \emph{Proceedings of the 2023 AAAI/ACM Conference on AI, Ethics, and Society}, 723--741.

\bibitem[{Slattery et~al.(2024)Slattery, Saeri, Grundy, Graham, Noetel, Uuk, Dao, Pour, Casper, and Thompson}]{slattery2024ai}
Slattery, P.; Saeri, A.~K.; Grundy, E.~A.; Graham, J.; Noetel, M.; Uuk, R.; Dao, J.; Pour, S.; Casper, S.; and Thompson, N. 2024.
\newblock The AI Risk Repository: A Comprehensive Meta-Review, Database, and Taxonomy of Risks From Artificial Intelligence.
\newblock \emph{AGI-Artificial General Intelligence-Robotics-Safety \& Alignment}, 1(1).

\bibitem[{Sokol et~al.(2024)Sokol, Moniz, Daly, Hind, and Chawla}]{sokol2024benchmarkcards}
Sokol, A.; Moniz, N.; Daly, E.; Hind, M.; and Chawla, N. 2024.
\newblock BenchmarkCards: Large Language Model and Risk Reporting.
\newblock \emph{arXiv preprint arXiv:2410.12974}.

\bibitem[{Stahl and Eke(2024)}]{stahl2024ethics}
Stahl, B.~C.; and Eke, D. 2024.
\newblock The ethics of ChatGPT--Exploring the ethical issues of an emerging technology.
\newblock \emph{International Journal of Information Management}, 74: 102700.

\bibitem[{Strachan et~al.(2024)Strachan, Albergo, Borghini, Pansardi, Scaliti, Gupta, Saxena, Rufo, Panzeri, Manzi et~al.}]{strachan2024testing}
Strachan, J.~W.; Albergo, D.; Borghini, G.; Pansardi, O.; Scaliti, E.; Gupta, S.; Saxena, K.; Rufo, A.; Panzeri, S.; Manzi, G.; et~al. 2024.
\newblock Testing theory of mind in large language models and humans.
\newblock \emph{Nature Human Behaviour}, 1--11.

\bibitem[{Tahaei et~al.(2023)Tahaei, Constantinides, Quercia, Kennedy, Muller, Stumpf, Liao, Baeza-Yates, Aroyo, Holbrook, Luger, Madaio, Blumenfeld, De-Arteaga, Vitak, and Olteanu}]{10.1145/3544549.3583178}
Tahaei, M.; Constantinides, M.; Quercia, D.; Kennedy, S.; Muller, M.; Stumpf, S.; Liao, Q.~V.; Baeza-Yates, R.; Aroyo, L.; Holbrook, J.; Luger, E.; Madaio, M.; Blumenfeld, I.~G.; De-Arteaga, M.; Vitak, J.; and Olteanu, A. 2023.
\newblock Human-Centered Responsible Artificial Intelligence: Current \& Future Trends.
\newblock In \emph{Extended Abstracts of the 2023 CHI Conference on Human Factors in Computing Systems}, CHI EA '23. New York, NY, USA: Association for Computing Machinery.
\newblock ISBN 9781450394222.

\bibitem[{Turri and Dzombak(2023)}]{turri2023we}
Turri, V.; and Dzombak, R. 2023.
\newblock Why we need to know more: Exploring the state of AI incident documentation practices.
\newblock In \emph{Proceedings of the 2023 AAAI/ACM Conference on AI, Ethics, and Society}, 576--583.

\bibitem[{Uuk et~al.(2024)Uuk, Brouwer, Schreier, Dreksler, Pulignano, and Bommasani}]{uuk2024effective}
Uuk, R.; Brouwer, A.; Schreier, T.; Dreksler, N.; Pulignano, V.; and Bommasani, R. 2024.
\newblock Effective Mitigations for Systemic Risks from General-Purpose AI.
\newblock \emph{arXiv preprint arXiv:2412.02145}.

\bibitem[{Vel{\'a}zquez et~al.(2024)Vel{\'a}zquez, {\v S}{\'c}epanovi{\'c}, Gvirtz, and Quercia}]{velazquezDecodingRealWorldArtificial2024}
Vel{\'a}zquez, J. D.~M.; {\v S}{\'c}epanovi{\'c}, S.; Gvirtz, A.; and Quercia, D. 2024.
\newblock Decoding {{Real-World Artificial Intelligence Incidents}}.
\newblock \emph{Computer}, 57(11): 71--81.

\bibitem[{Wang et~al.(2021)Wang, Zhu, Liu, and Sun}]{Wang2021}
Wang, Z.; Zhu, H.; Liu, P.; and Sun, L. 2021.
\newblock Social engineering in cybersecurity: a domain ontology and knowledge graph application examples.
\newblock \emph{Cybersecurity}, 4(1): 31.

\bibitem[{Wang et~al.(2024)Wang, Kulkarni, Wilcox, Terry, and Madaio}]{wang2024farsight}
Wang, Z.~J.; Kulkarni, C.; Wilcox, L.; Terry, M.; and Madaio, M. 2024.
\newblock Farsight: Fostering Responsible AI Awareness During AI Application Prototyping.
\newblock In \emph{Proceedings of the CHI Conference on Human Factors in Computing Systems}, 1--40.

\bibitem[{Washo(2021)}]{WASHO2021100126}
Washo, A.~H. 2021.
\newblock An interdisciplinary view of social engineering: A call to action for research.
\newblock \emph{Computers in Human Behavior Reports}, 4: 100126.

\bibitem[{Weidinger et~al.(2021)Weidinger, Mellor, Rauh, Griffin, Uesato, Huang, Cheng, Glaese, Balle, Kasirzadeh et~al.}]{weidinger2021ethical}
Weidinger, L.; Mellor, J.; Rauh, M.; Griffin, C.; Uesato, J.; Huang, P.-S.; Cheng, M.; Glaese, M.; Balle, B.; Kasirzadeh, A.; et~al. 2021.
\newblock Ethical and social risks of harm from language models.
\newblock \emph{arXiv preprint arXiv:2112.04359}.

\bibitem[{Weidinger et~al.(2023)Weidinger, Rauh, Marchal, Manzini, Hendricks, Mateos-Garcia, Bergman, Kay, Griffin, Bariach et~al.}]{weidinger2023sociotechnical}
Weidinger, L.; Rauh, M.; Marchal, N.; Manzini, A.; Hendricks, L.~A.; Mateos-Garcia, J.; Bergman, S.; Kay, J.; Griffin, C.; Bariach, B.; et~al. 2023.
\newblock Sociotechnical safety evaluation of generative ai systems.
\newblock \emph{arXiv preprint arXiv:2310.11986}.

\bibitem[{Weidinger et~al.(2022)Weidinger, Uesato, Rauh, Griffin, Huang, Mellor, Glaese, Cheng, Balle, Kasirzadeh, Biles, Brown, Kenton, Hawkins, Stepleton, Birhane, Hendricks, Rimell, Isaac, Haas, Legassick, Irving, and Gabriel}]{weidinger2022taxonomy}
Weidinger, L.; Uesato, J.; Rauh, M.; Griffin, C.; Huang, P.-S.; Mellor, J.; Glaese, A.; Cheng, M.; Balle, B.; Kasirzadeh, A.; Biles, C.; Brown, S.; Kenton, Z.; Hawkins, W.; Stepleton, T.; Birhane, A.; Hendricks, L.~A.; Rimell, L.; Isaac, W.; Haas, J.; Legassick, S.; Irving, G.; and Gabriel, I. 2022.
\newblock Taxonomy of Risks posed by Language Models.
\newblock In \emph{Proceedings of the 2022 ACM Conference on Fairness, Accountability, and Transparency}, FAccT '22, 214–229. New York, NY, USA: Association for Computing Machinery.
\newblock ISBN 9781450393522.

\bibitem[{Whittaker et~al.(2019)Whittaker, Alper, Bennett, Hendren, Kaziunas, Mills, Morris, Rankin, Rogers, Salas et~al.}]{whittaker2019disability}
Whittaker, M.; Alper, M.; Bennett, C.~L.; Hendren, S.; Kaziunas, L.; Mills, M.; Morris, M.~R.; Rankin, J.; Rogers, E.; Salas, M.; et~al. 2019.
\newblock Disability, bias, and {AI}.
\newblock \emph{AI Now Institute}, 8.

\bibitem[{Wirtz, Weyerer, and Sturm(2020)}]{wirtz2020dark}
Wirtz, B.~W.; Weyerer, J.~C.; and Sturm, B.~J. 2020.
\newblock The dark sides of artificial intelligence: An integrated AI governance framework for public administration.
\newblock \emph{International Journal of Public Administration}, 43(9): 818--829.

\bibitem[{Yampolskiy(2016)}]{yampolskiy2016taxonomy}
Yampolskiy, R.~V. 2016.
\newblock Taxonomy of pathways to dangerous artificial intelligence.
\newblock In \emph{Workshops at the thirtieth AAAI conference on artificial intelligence}.

\bibitem[{Yin et~al.(2024)Yin, Fu, Zhao, Li, Sun, Xu, and Chen}]{10.1093/nsr/nwae403}
Yin, S.; Fu, C.; Zhao, S.; Li, K.; Sun, X.; Xu, T.; and Chen, E. 2024.
\newblock A survey on multimodal large language models.
\newblock \emph{National Science Review}, 11(12): nwae403.

\bibitem[{Yu et~al.(2024)Yu, Zhuang, Zhang, Meng, Ratner, Krishna, Shen, and Zhang}]{yu2024large}
Yu, Y.; Zhuang, Y.; Zhang, J.; Meng, Y.; Ratner, A.~J.; Krishna, R.; Shen, J.; and Zhang, C. 2024.
\newblock Large language model as attributed training data generator: A tale of diversity and bias.
\newblock \emph{Advances in Neural Information Processing Systems}, 36.

\bibitem[{Ziems et~al.(2024)Ziems, Held, Shaikh, Chen, Zhang, and Yang}]{ziems2024can}
Ziems, C.; Held, W.; Shaikh, O.; Chen, J.; Zhang, Z.; and Yang, D. 2024.
\newblock Can large language models transform computational social science?
\newblock \emph{Computational Linguistics}, 50(1): 237--291.

\end{thebibliography}
